\documentclass[toc]{PoS}
\pdfoutput=1

\usepackage{amsmath,amssymb,amsfonts,bm} 
\usepackage{subfigure}
\usepackage{mathtools}
\usepackage{etex}
\usepackage{tikz}
\usepackage{tkz-euclide}
\usepackage{cite}
\usepackage{feynmp}
\usetkzobj{all}
\usetikzlibrary{matrix,positioning,arrows}

\DeclareMathOperator{\diag}{diag}
\DeclareMathOperator{\imag}{Im}

\DeclareMathOperator{\tr}{Tr}

\newtheorem{exercise}{Exercise}
\newtheorem{example}{Example}

\DeclareFontFamily{OMS}{rsfs}{\skewchar\font'177}
\DeclareFontShape{OMS}{rsfs}{m}{n}{%
      <5> rsfs5 <6> <7> rsfs7 <8> <9> <10> rsfs10
      <10.95> <12> <14.4> <17.28> <20.74> <24.88> rsfs10}{}
\DeclareSymbolFont{rsfscript}{OMS}{rsfs}{m}{n}
\DeclareSymbolFontAlphabet{\mathrsfs}{rsfscript}

\newcommand{\Lag}{\mathrsfs{L}}
\def\cp{\ensuremath{(\text{CP})\,}}
\def\invcp{\ensuremath{\,(\text{CP})}^{-1}}
\newcommand{\sempty}[1]{}

\usetikzlibrary{trees,shapes,decorations.markings}
\usetikzlibrary{positioning,arrows}
\usetikzlibrary{decorations.pathmorphing}
\tikzset{
  gauge/.style={decorate, decoration={snake}},
  scalar/.style={dashed},
  fermion/.style={postaction={decorate},
    decoration={markings,mark=at position .55 with {\arrow{>}}}},
  gluon/.style={decorate, decoration={coil,amplitude=4pt, segment length=5pt}} 
}
 
\title{Flavour Physics and CP Violation in the Standard Model and
Beyond}

\ShortTitle{Flavour Physics and CP Violation in the SM and Beyond}

\author{Gustavo Castelo-Branco\\ CFTP-Instituto Superior T\'ecnico\\
Universidade T\'ecnica de Lisboa\\ Av. Rovisco Pais, 1049-001 Lisboa,
Portugal\\ E-mail: \email{gbranco@tecnico.ulisboa.pt}}

\author{David Emmanuel-Costa\\ IST-ID and CFTP-Instituto Superior
T\'ecnico\\ Universidade T\'ecnica de Lisboa\\ Av. Rovisco Pais,
1049-001 Lisboa, Portugal\\ E-mail: \email{david.costa@tecnico.ulisboa.pt}}
\abstract{}

\FullConference{\textbf{Third IDPASC School}\\ San Marti\~no Pinario\\
Santiago de Compostela\\ Galiza, Spain\\ January 21st - February 2nd
2013}

\begin{document}

\section{Introduction}

 We present the invited lectures given at the Third IDPASC School which
took place in Santiago de Compostela in January 2013. The students
attending the school had very different backgrounds, some of them were
doing their PhD in experimental particle physics, others in theory. As
a result, and in order to make the lectures useful for most of the
students, we focused on basic topics of broad interest, avoiding the
more technical aspects of Flavour Physics and CP Violation.  We make a
brief review of the Standard Model, paying special attention to the
generation of fermion masses and mixing, as well as to CP violation.
We describe some of the simplest extensions of the SM, emphasising
novel flavour aspects which arise in their framework.

\section{Review of the Standard Model}

The Standard Model (SM) of unification of the electroweak and strong
interactions~\cite{Glashow:1961tr,Weinberg:1967tq,Salam:1968rm,Glashow:1970gm} is
based on the gauge group
\begin{equation} 
G_{\text{SM}}\equiv\mathsf{SU}(3)_{\text{\sc c}}
\times \mathsf{SU}(2)_{\text{\sc l}} \times \mathsf{U}(1)_{\text{\sc
y}}\,,
\end{equation} which has 12 generators. To each one of these
generators corresponds a gauge field. The introduction of these gauge
fields is essential in order to achieve invariance under local gauge
transformations of $G_{\text{SM}}$.  This is entirely analogous to
what one encounters in electromagnetic interactions, where the photon
is the gauge field associated to the $\mathsf{U}(1)_{\text{e.m.}}$,
introduced in order to guarantee local gauge invariance. We shall
denote the gauge fields in the following way:
\begin{align} \mathsf{SU}(3)_{\text{\sc c}} \,&\longrightarrow\;
G_{\mu}^{k},\;k=1,\dots,8\,;\\ \mathsf{SU}(2)_{\text{\sc l}}
\,&\longrightarrow\; W_{\mu}^{j},\;j=1,\dots,3\,;\\
\mathsf{SU}(2)_{\text{\sc l}} \,&\longrightarrow\; B_{\mu}\,.
\end{align} The electroweak interactions are linear combination of the
following gauge bosons:
\begin{equation}
W_{\mu}^{a},\,B_{\mu}\;\longrightarrow\;W^{+}_{\mu},\,W^{-}_{\mu},\,Z_{\mu},\,A_{\mu}\,,
\end{equation} where $A_{\mu}$ is the photon field, mediator of
electromagnetic interactions while the massive bosons $W^{+}_{\mu}$ and
$Z_{\mu}$ mediate, respectively, the charged and neutral weak
currents. Since $\mathsf{U}(1)_{\text{e.m.}}$ is a good symmetry of
nature, the photon field should remain massless.

The SM describes all observed fermionic particles, which have definite
gauge transformations properties and are replicated in three
generations. All the SM fermionic fields carry weak hypercharge $Y$,
defined as
\begin{equation}
\label{eq:ydef} Y\,\equiv\,Q-T_3\,,
\end{equation} where $Q$ is the electric charge operator and $T_3$ is
the diagonal generator of $\mathsf{SU}(2)_{\text{\sc l}}$. Since
experiments only provided evidence for left-handed charged currents, the
right-handed components of fermion fields are put in
$\mathsf{SU}(2)_{\text{\sc l}}$-singlets. Only the quarks carry
colour, i.e they are triplets of $\mathsf{SU}(3)_{\text{\sc c}}$,
while the leptons carry no colour. We summarise in Table~\ref{tab:SM}
all fermionic content characterised by their transformation properties
under the gauge group $\mathsf{SU}(3)_{\text{\sc c}}\,\times\,
\mathsf{SU}(2)_{\text{\sc l}}\,\times\,\mathsf{U}(1)_Y$. It is worth
noting that within this matter content the SM is free from anomalies,
since $\mathsf{SU}(3)_{\text{\sc c}}$ is non-chiral, all
representations of $\times \mathsf{SU}(2)_{\text{\sc l}}$ are real,
the $\mathsf{SU}(3)^2\,Y$, $\mathsf{SU}(2)^2\,Y$ and $Y^3$ cancel
between the quarks and leptons. 
 
\begin{table}[t]
  \caption{\label{tab:SM} The SM fermionic content. For a given SM
representation~$R$ one has
$(n_3,n_2,y)\equiv(\dim_{\mathsf{SU}(3)}(R),\dim_{\mathsf{SU}(2)}(R),Y(R))\,.$
The index $i=1,2,3$ is the generation index.}
\begin{center}
\begin{tabular}{lc} \hline\hline \\
${q_i}_L\equiv\begin{pmatrix}u_i\\d_i\end{pmatrix}_L$ &
$(3,2,1/6)$\\[4mm] ${u_i}_R$ & $(3,1,2/3)$\\[4mm] ${d_i}_R$ &
$(3,1,-1/3)$\\[4mm]
${\ell_i}_L\equiv\begin{pmatrix}\nu_i\\e^{-}_i\end{pmatrix}_{\text{\sc
l}}$ & $(1,2,-1/2)$\\[4mm] ${e^{-}_i}_R$ & $(1,1,-1)$\\ \\
\hline\hline
\end{tabular}
\end{center}
\end{table}

Gauge interactions are determined by the covariant derivative which is
dictated by the transformation properties of the various fields, under
the gauge group. In general one has
\begin{equation}
D_{\mu}\,=\,\partial_{\mu}-ig_sL^kG^k_{\mu}-igT^jW^j_{\mu}-ig'y\,B_{\mu}\,,
\end{equation} 
where $T^j$ are the three $\mathsf{SU}(2)$-generators,
\begin{equation} 
T^j\,=\,\left\{
\begin{array}{cc}
0\,, & \text{singlet} \\ 
\frac{\tau_j}{2}\,,& \text{fundamental}
\end{array}
\right.\,,
\end{equation} 
while $L^k$ are the eight $\mathsf{SU}(3)$-generators,
\begin{equation} 
L^k\,=\,\left\{
\begin{array}{cc}
0\,, & \text{singlet}\\
\frac{\lambda_k}{2}\,,& \text{fundamental}
\end{array}
\right.\,.
\end{equation} 
The matrices $\tau_j$ and $\lambda_k$ are the usual
Pauli and Gell-Mann matrices, respectively. For the fermions presented
in Table~\ref{tab:SM} the covariant derivatives read as
\begin{align}
D_{\mu}\,q_L&=\left(\partial_{\mu}-i\frac{g_s}{2}\lambda_k\,G^k_{\mu}-i\frac{g}{2}\tau_j\,W^j_{\mu}-i\frac{g'}6B_{\mu}\right)\,q_L\,,\\
D_{\mu}\,u_R&=\left(\partial_{\mu}-i\frac{g_s}{2}\lambda_k\,G^k_{\mu}-i\frac{2g'}3B_{\mu}\right)\,u_R\,,\\
D_{\mu}\,d_R&=\left(\partial_{\mu}-i\frac{g_s}{2}\lambda_k\,G^k_{\mu}+i\frac{g'}3B_{\mu}\right)\,d_R\,,\\
D_{\mu}\,\ell_L&=\left(\partial_{\mu}-i\frac{g}{2}\tau_j\,W^j_{\mu}+i\frac{g'}2B_{\mu}\right)\,q_L\,,\\
D_{\mu}\,e^{-}_L&=\left(\partial_{\mu}+igB_{\mu}\right)\,e^{-}_R\,.
\end{align}

An important feature of the SM is the fact that right-handed neutrinos,
\begin{equation}
\nu_R\,\sim\,(1,1,0)\,,
\end{equation}
are not introduced. As a result, neutrinos are strictly massless in the
SM, in contradiction with present experimental evidence. We shall come
back to this question in the sequel.

\begin{figure}[t]
  \begin{center} \subfigure[f1][$\lambda>0$,\,
$\mu^2>0$]{\includegraphics[width=7cm]{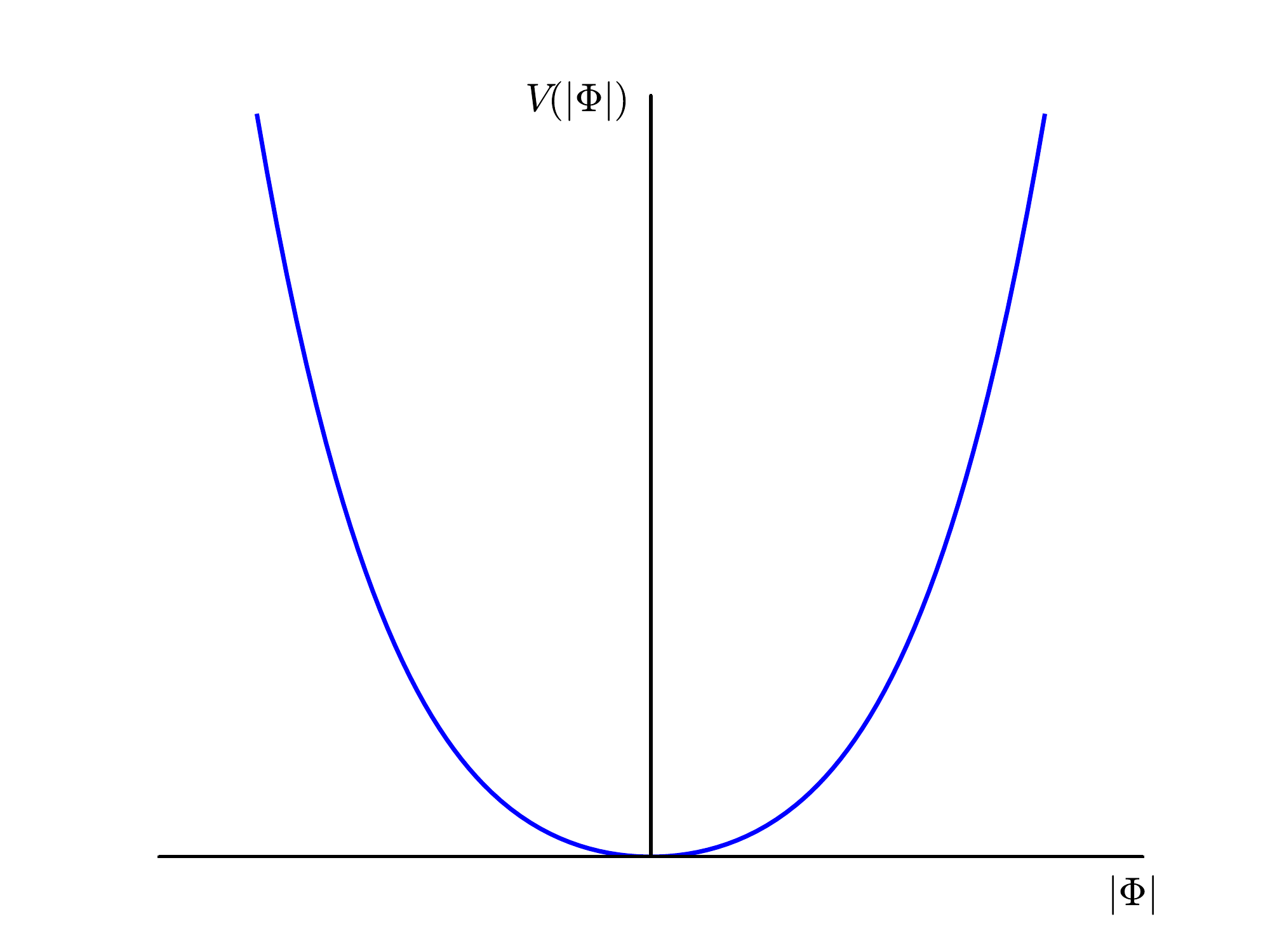}} \qquad \subfigure[f2][$\lambda>0$,\,
$\mu^2<0$]{\includegraphics[width=7cm]{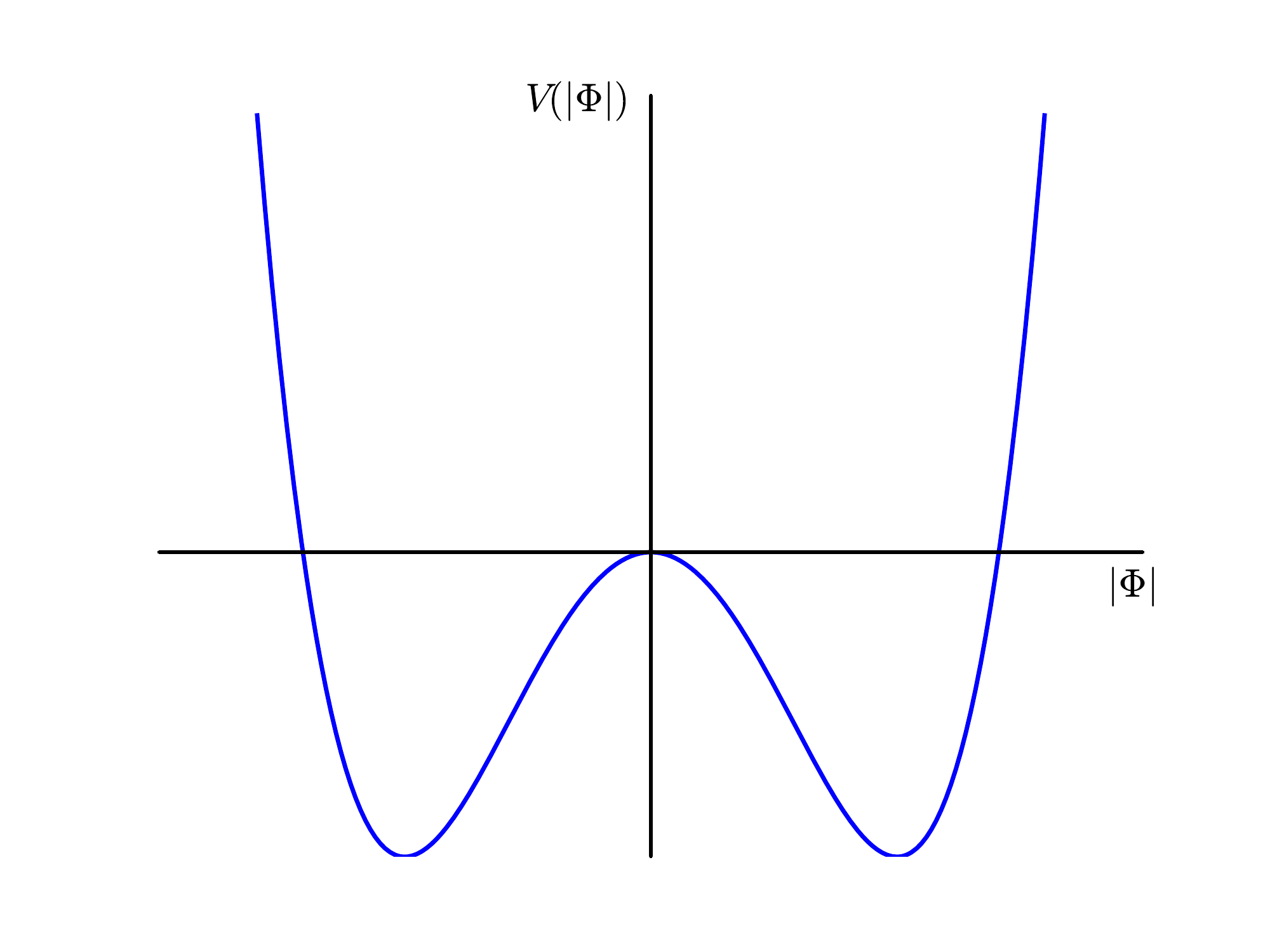}}
  \end{center}
\caption{\label{fig:potential}}
\end{figure}

In order to account for the massive gauge bosons $W^{\pm}_{\mu}$ and
$Z_{\mu}$ without destroying renormalisability, the gauge symmetry
must be spontaneously broken. The simplest scheme to break
spontaneously the electroweak gauge symmetry into electromagnetism,
involves the introduction of a complex doublet Higgs scalar field
$\phi$
\begin{equation} \phi\,=\,\begin{pmatrix}\phi^{+} \\ \phi^0
\end{pmatrix}\;\sim\;(1,2,1/2)\,,
\end{equation} 
which leads to the breaking:
\begin{equation}
  \label{eq:bs} \mathsf{SU}(3)_{\text{\sc c}} \times
\mathsf{SU}(2)_{\text{\sc l}} \times \mathsf{U}(1)_{\text{\sc y}}
\;\longrightarrow\; \mathsf{SU}(3)_{\text{\sc c}} \times
\mathsf{U}(1)_{\text{e.m.}}\,.
\end{equation} 
The most general gauge invariant, renormalisable scalar potential is:
\begin{equation}
  \label{eq:V}
V(\phi)=\mu^2\phi^{\dagger}\phi\,+\,\lambda\left(\phi^{\dagger}\phi\right)^2\,.
\end{equation} Taking $\lambda>0$ the potential is bounded from below
and two minima do exist. For $\mu^2>0$ one has
$\langle0|\phi|0\rangle=0$ while for $\mu^2<0$ one has instead
\begin{equation}
  \label{eq:VEV}
\langle0|\phi|0\rangle\,=\,\begin{pmatrix}0\\\frac{1}{\sqrt{2}}v\end{pmatrix}\,.
\end{equation} 
In Figure~\ref{fig:potential} it is drawn the Higgs
potential around the two minima. Indeed, the case $\lambda>0$ and
$\mu^2<0$ implies the spontaneous breaking of the electroweak gauge
 as indicated in eq.~\ref{eq:bs}.  One
can check that the $\mathsf{U}(1)$ remains unbroken. The
electric charge operator reads as
\begin{equation}
  \label{eq:Q} Q\,=\,T_3\,+\,Y\,,
\end{equation} and for the Higgs doublet one gets
\begin{equation}
  \label{eq:Q1} Q\,=\,\begin{pmatrix} \frac12 & \ 0\ \\ \ 0\ &
-\frac12 \end{pmatrix}\,+\,
\begin{pmatrix}\frac12 & \ 0\ \\\ 0\ & \frac12 \end{pmatrix}
\,=\,\begin{pmatrix}\ 1\ &\ 0\ \\ 0 & 0 \end{pmatrix}\,.
\end{equation} 
Therefore one verifies that the vacuum given in
eq.~\eqref{eq:VEV} is invariant under the charge operator $Q$, since
\begin{equation}
  \label{eq:Q2}
Q\,\begin{pmatrix}0\\\frac{1}{\sqrt{2}}v\end{pmatrix}\,=\,0\,,
\end{equation} and one gets
\begin{equation}
  \label{eq:em}
  e^{i\,\alpha\,Q}\, 
\begin{pmatrix}0\\\frac{1}{\sqrt{2}}v\end{pmatrix} 
\,=\, \bigg[1\,+\,i\,\alpha\,Q\,+\,\cdots\bigg]\, 
\begin{pmatrix}0\\\frac{1}{\sqrt{2}}v\end{pmatrix}
\,=\,\begin{pmatrix}0\\\frac{1}{\sqrt{2}}v\end{pmatrix}\,.
\end{equation}
Electric charge is automatically conserved in the SM. This is no
longer true in extensions of the SM with two Higgs doublets, 
including the case of supersymmetric extensions of the SM. In the general two
Higgs doublet model (2HDM) without loss of
generality, one has:
\begin{equation}
  \label{eq:VEVDH}
\langle0|\phi_{1}|0\rangle\,=\,\begin{pmatrix}0\\\frac{1}{\sqrt{2}}v_{1}\end{pmatrix}\,,
\qquad
\langle0|\phi_{2}|0\rangle\,=\,\begin{pmatrix}\xi\\\frac{1}{\sqrt{2}}v_{2}\,e^{i\,\theta}\end{pmatrix}\,,
\end{equation}
with $\xi$ real. In order to preserve charge conservation in the 2HDM, one
has to choose a region of the parameter space where the minimum is at
$\xi=0\,$.

The SM does not provide an explanation for the charges of elementary
fermions. The values of the hypercharge $Y$ are chosen in such a way
that the correct electric charges are obtained. As an example, one can
determined $Y_{q_{L}}\,$, by using the eq.~\eqref{eq:ydef} and the
knowledge of $Q_{u}$ and $Q_{d}\,$. Thus,
\begin{align} Y_{u_{L}}&=\,\frac23-\frac12\,=\,\frac16\,,\\
Y_{d_{L}}&=\,-\frac13+\frac12\,=\,\frac16\,,
\end{align} and therefore $Y_{q_{L}}=1/6\,$. It is rather intriguing
the fact that the requirement of cancelation of the gauge anomaly in the SM
together with the fact that the electromagnetic interactions are
non-chiral is sufficient to fully determine all the hypercharges of
the fundamental fermions up to an overall factor. In particular one gets relations
among quark and lepton charges, leading to:
\begin{equation}
  \label{eq:Charge} Q_{p}\,=\,-Q_{e}\,.
\end{equation} 
\emph{Although the hypercharge quantisation may arise from the
anomaly-free condition, this is certainly not a satisfactory
explanation in the SM. The solution to this fundamental question is
elegantly answered in the framework of Grand-Unification,
e.g. $\mathsf{SU}(5)$, where the quantisation of electric charges
is related to some new phenomena like the magnetic monopoles
predicted in the theory that can be tested in future experiments.}

  In order to describe the spontaneous breaking of the electroweak
symmetry in the SM, one starts by introducing a convenient
parametrisation of the Higgs doublet $\phi$ as
\begin{equation}
  \label{eq:phi}
\phi\,=\,\begin{pmatrix}G^{+}\\\frac1{\sqrt{2}}(v\,+\,H\,+\,i\,G_{0})\end{pmatrix}\,,
\end{equation} 
where $G^{+}$ is a charged complex scalar field, $H$ is
a real scalar field and $G_{0}$ is a real pseudo-scalar. The scalar
fields $G^{\pm}$ and $G_{0}$ are massless states, the so-called
Nambu-Goldstone bosons. Through the Brout-Englert-Higgs mechanism, the charged bosons $G^{\pm}$ are absorbed as
longitudinal components of the $W^{\pm}_{\mu}$ which acquire a mass:
\begin{equation}
  \label{eq:MW} M_{W}\,=\,\frac{g\,v}{2}\,,
\end{equation} while the neutral boson $G_{0}$ becomes the
longitudinal component of the gauge boson $Z_{\mu}$, which is a linear
combination of the bosons $B_{\mu}$ and $W^{3}_{\mu}$,
\begin{equation}
  \label{eq:Z}
Z_{\mu}\,=\,\cos\theta_{W}\,W^{3}_{\mu}\,-\,\sin\theta_{W}\,B_{\mu}\,,
\end{equation} where $\theta_{W}$ is simply given by
\begin{equation}
  \label{eq:thetaW} \tan\theta_{W}\,\equiv\,\frac{g'}{g}\,.
\end{equation} The $Z_{\mu}$ boson acquires then a mass given by
\begin{equation}
  \label{eq:MZ}
M_{Z}\,=\,\sqrt{g^{2}\,+\,g'^{2}}\,\frac{v}{2}\,=\,\frac{M_{W}}{\cos\theta_{W}}\,.
\end{equation} 
The bosonic state orthogonal bosonic state to $Z_{\mu}$:
\begin{equation}
  \label{eq:gg}
A_{\mu}\,=\,\cos\theta_{W}\,B_{\mu}\,+\,\sin\theta_{W}\,W^{3}_{\mu}\,,
\end{equation}
remains massless and is identified with the photon. The electron
coupling to the photon is directly determined from the weak couplings $g$ and $g'$ as
\begin{equation}
\label{rel:eggp}
\frac1{e^2}\,=\,\frac1{g^2}+\frac1{g'^2}\,,
\end{equation}
or
\begin{equation}
e=\frac{g\,g'}{\sqrt{g^2+g'^2}}\,=\,g\sin\theta_W\,=\,g'\cos\theta_W\,.
\end{equation}

\section{Fermion masses and mixings}

In the SM, one cannot write directly a mass term for any of the fundamental
fermions because they would violate the gauge symmetry, since
left-handed and right-handed chiralities do transform differently. The
SM fermions acquire mass through Yukawa couplings, once the SM
group is spontaneously broken. Therefore, in the SM the
Higgs mechanism that is responsible for the breaking of the SM group, also
generates fermion masses.

\subsection*{Quark and Charged Lepton masses}

The Yukawa  interactions are the most general terms in the
Lagrangian allowed by the SM gauge group that involve fermions and the
Higgs doublet. The Yukawa couplings can be written as:
\begin{equation}
  \label{eq:Ly}
  -\Lag_Y \,=\, \left(Y_u\right)_{ij}\,\overline{q_i}_L\,\tilde{\phi}\,{u_i}_R
   \,+\,\left(Y_d\right)_{ij}\,\overline{q_i}_L\,\phi\,{d_i}_R
   \,+\,\left(Y_{\ell}\right)_{ij}\,\overline{\ell_i}_L\,\phi\,{e_i}_R
  \,+\,\text{H.c.}\,,
\end{equation}
where $\tilde{\phi}\equiv\,i\tau_2\,\phi^{\dagger}$. The Yukawa
matrices $Y_u$, $Y_d$ and $Y_{\ell}$ are arbitrary complex matrices in
flavour space. The first two terms in eq.~\eqref{eq:Ly} will
generate  the up- and down-type quark masses while the third term will
give rise to the charged lepton masses. Making use of the Higgs doublet
parametrisation given in eq.~\eqref{eq:phi} one can decompose the
Lagrangian given in eq.~\eqref{eq:Ly} as
\begin{equation}
  \label{eq:Lyy}
  \begin{split}
    -\Lag_Y \,
    &=\,   \frac{v}{\sqrt2}\,\left(Y_u\right)_{ij}\,\overline{u_i}_L\,{u_i}_R
    \,+\,  \frac{v}{\sqrt2}\,\left(Y_d\right)_{ij}\,\overline{d_i}_L\,{d_i}_R
    \,+\,  \frac{v}{\sqrt2}\,\left(Y_{\ell}\right)_{ij}\,\overline{e_i}_L\,{e_i}_R
    \\
    &\,+\,\frac{\left(Y_u\right)_{ij}}{\sqrt2}\,\overline{u_i}_L\,{u_i}_R\,H
    \,+\,  \frac{\left(Y_d\right)_{ij}}{\sqrt2}\,\overline{d_i}_L\,{d_i}_R\,H
    \,+\,  \frac{\left(Y_{\ell}\right)_{ij}}{\sqrt2}\,\overline{e_i}_L\,{e_i}_R\,H
    \\
    &\,-\,\frac{i\left(Y_u\right)_{ij}}{\sqrt2}\,\overline{u_i}_L\,{u_i}_R\,G^0
    \,+\,  \frac{i\left(Y_d\right)_{ij}}{\sqrt2}\,\overline{d_i}_L\,{d_i}_R\,G^0
    \,+\,  \frac{i\left(Y_{\ell}\right)_{ij}}{\sqrt2}\,\overline{e_i}_L\,{e_i}_R\,G^0
    \\
    &\,-\,\left(Y_u\right)_{ij}\,\overline{d_i}_L\,{u_i}_R\,G^{-}
    \,+\,\left(Y_d\right)_{ij}\,\overline{u_i}_L\,{d_i}_R,\,G^{+}
    \,+\,\left(Y_{\ell}\right)_{ij}\,\overline{\nu_i}_L\,{e_i}_R\,G^{+}
    \,+\,\text{H.c.}\,.
  \end{split}
\end{equation}
Once a gauge transformation is performed in order to absorbed the
Nambu-Goldstone bosons $G^{\pm}$ and $G^0$, the Lagrangian in
eq.~\eqref{eq:Lyy} becomes
\begin{equation}
  \label{eq:Lyyy}
  \begin{split}
    -\Lag_Y \,
    &=\,  \left(m_u\right)_{ij}\,\overline{u_i}_L\,{u_i}_R
    \,+\, \left(m_d\right)_{ij}\,\overline{d_i}_L\,{d_i}_R
    \,+\,  \left(m_{\ell}\right)_{ij}\,\overline{e_i}_L\,{e_i}_R
    \\
    &\,+\,\frac{\left(Y_u\right)_{ij}}{\sqrt2}\,\overline{u_i}_L\,{u_i}_R\,H
    \,+\,  \frac{\left(Y_d\right)_{ij}}{\sqrt2}\,\overline{d_i}_L\,{d_i}_R\,H
    \,+\,  \frac{\left(Y_{\ell}\right)_{ij}}{\sqrt2}\,\overline{e_i}_L\,{e_i}_R\,H
    \,+\,\text{H.c.}\,,
  \end{split}
\end{equation}
where the quark mass matrices $m_u$, $m_d$ and the charged lepton mass
matrix $m_{\ell}$ are simply defined by
\begin{equation}
  \label{eq:mdeff}
  m_u\,\equiv\,\frac{v}{\sqrt2}\,Y_u\,,\qquad
  m_d\,\equiv\,\frac{v}{\sqrt2}\,Y_d\,,\qquad
  m_{\ell}\,\equiv\,\frac{v}{\sqrt2}\,Y_{\ell}\,.
\end{equation}
Gauge invariance does not constrain the flavour structure of Yukawa
couplings and therefore $m_u$, $m_d$ and $m_{\ell}$ are arbitrary complex
matrices.

Let us now focus on the mass terms,
\begin{equation}
  \label{eq:Lm}
    -\Lag_m 
    \,=\,  \left(m_u\right)_{ij}\,{\overline{u_i}}^0_L\,{u^0_i}_R
     \,+\, \left(m_d\right)_{ij}\,{\overline{d_i}}^0_L\,{d^0_i}_R
    \,+\,  \left(m_{\ell}\right)_{ij}\,{\overline{e_i}}^0_L\,{e^0_i}_R\,.
\end{equation}
A super-script $0$ on the fermion fields was used that these fields are the original ones, in the
weak basis. The matrices $m_{u,d,e}$ can be diagonalised by the
following bi-unitary transformations:
\begin{subequations}
\label{eq:diagU}
\begin{align}
{u_i}^0_L\,&=\,U_L^u\,{u_i}_L\,;\qquad {u_i}^0_R\,=\,U_R^u\,{u_i}_R\,,\\
{d_i}^0_L\,&=\,U_L^d\,{d_i}_L\,;\qquad {d_i}^0_R\,=\,U_R^d\,{d_i}_R\,,\\
{e_i}^0_L\,&=\,U_L^e\,{e_i}_L\,;\qquad {e_i}^0_R\,=\,U_R^e\,{e_i}_R\,,
\end{align}
\end{subequations}
where $U^{u,d,e}_{R,L}$ are a set of unitary matrix such as
\begin{subequations}
\begin{align}
m_u\,\longrightarrow\,{U_L^u}^{\dagger}\,m_u\,U_R^u\,&=\,\diag\left(m_u,m_c,m_t\right)\,,\\
m_d\,\longrightarrow\,{U_L^d}^{\dagger}\,m_d\,U_R^d\,&=\,\diag\left(m_d,m_s,m_b\right)\,,\\
m_{\ell}\,\longrightarrow\,{U_L^e}^{\dagger}\,m_{\ell}\,U_R^e\,&=\,\diag\left(m_{\ell},m_{\mu},m_{\tau}\right)\,.
\end{align}
\end{subequations}
The fields $u_{L,R},d_{L,R},e_{L,R}$ are thus the mass
eigenstates. The bi-unitary transformations given in
eq.~\eqref{eq:diagU} affect the interactions between left-handed
particles and the $W^{\pm}_{\mu}$ bosons - the \emph{charged currents}
- which are written in a weak basis as:
\begin{equation}
  \label{eq:Lcc}
    -\Lag_{\text{\sc cc}} 
    \,=\, \frac{g}{\sqrt{2}}\left[ {\overline{u_i}}^0_L\,\gamma^{\mu}\,{d_i}^0_L 
      \,+\,
      {\overline{\nu_i}}^0_L\,\gamma^{\mu}\,{e_i}^0_L\right]\,W^{+}_{\mu}\,+\,\text{H.c.}\,.
\end{equation}
In the mass eigenstate basis the charged currents become:
\begin{equation}
  \label{eq:Lccfb}
    -\Lag_{\text{\sc cc}} 
    \,=\, \frac{g}{\sqrt{2}}\left[ \overline{u}_L\,\gamma^{\mu}\,{U_L^u}^{\dagger}U_L^d\,d_L 
      \,+\,
      {\overline{\nu}}^0_L\,\gamma^{\mu}\,U_L^e\,e_L\right]\,W^{+}_{\mu}\,+\,\text{H.c.}\,.
\end{equation}
The product of unitary matrices in  eq.~\eqref{eq:Lccfb}
defines the well know Cabibbo-Kobayshi-Maskawa matrix $V$ as
 \begin{equation}
\label{eq:CKM}
V\,\equiv\,{U_L^u}^{\dagger}U_L^d\,.
 \end{equation}
 In the SM the unitary matrix $U_L^e$ is physically meaningless. Note
 that since neutrinos are massless in the SM, one can always redefine
 neutrino fields as
\begin{equation}
\nu^0_L\,\longrightarrow\,\nu_L\,=\,U_L^e\,{\nu}_L\,,
\end{equation}
and therefore the charged current term $
{\overline{\nu}}^0_L\,\gamma^{\mu}\,U_L^e\,e_L$ in
eq.~\eqref{eq:Lccfb} becomes $
{\overline{\nu}}_L\,\gamma^{\mu}\,e_L\,$. We then conclude that in the
SM there is no leptonic mixing and therefore no neutrino oscillations.

We can show that the electromagnetic and neutral currents are not
affected by the transformations given in eq.~\eqref{eq:diagU}.  The
electromagnetic $J_{\text{e.m.}}$ given in the weak basis,
\begin{equation}
\label{eq:Jem}
J^{\mu}_{\text{e.m.}}\,=\,\frac23\,\left[\,\overline{u}^0_L\,\gamma^{\mu}\,u^0_L\,+\,\overline{u}^0_R\,\gamma^{\mu}\,u^0_R\,\right]
\,-\frac13\,\left[\,\overline{d}^0_L\,\gamma^{\mu}\,d^0_L\,+\,\overline{d}^0_R\,\gamma^{\mu}\,d^0_R\,\right]
\,-\,\left[\,\overline{e}^0_L\,\gamma^{\mu}\,e^0_L\,+\,\overline{e}^0_R\,\gamma^{\mu}\,e^0_R\,\right]\,,
\end{equation}
do not change in the mass eigenstate, since $J^{\mu}_{\text{e.m.}}$ transforms as
\begin{equation}
\begin{split}
J'^{\mu}_{\text{e.m.}}\,=&\,\frac23\,\left[\,\overline{u}_L\,\gamma^{\mu}\,{U_L^u}^{\dagger}U_L^u\,u_L\,+\,\overline{u}_R\,\gamma^{\mu}\,{U_R^u}^{\dagger}U_R^u\,u_R\,\right]
\,-\frac13\,\left[\,\overline{d}_L\,\gamma^{\mu}\,{U_L^d}^{\dagger}U_L^d\,d_L\,+\,\overline{d}_R\,\gamma^{\mu}\,{U_R^d}^{\dagger}U_R^d\,d_R\,\right]\\
\,-\,&\left[\,\overline{e}_L\,\gamma^{\mu}\,{U_L^e}^{\dagger}U_L^e\,e_L\,+\,\overline{e}_R\,\gamma^{\mu}\,{U_R^e}^{\dagger}U_R^e\,e_R\,\right]\,,
\end{split}
\end{equation}
and we get the same formal expression as in eq.~\eqref{eq:Jem}.
In a similar way we demonstrate that the neutral currents Lagrangian, 
\begin{equation}
\Lag_{\text{\sc NC}}\,=\,\frac{g}{\cos\theta_W}\,\left[\,\overline{u}^0_L\,\gamma^{\mu}\,u^0_L\,-\,\overline{d}^0_L\,\gamma^{\mu}\,d^0_L
\,+\,
\,\overline{\nu}^0_L\,\gamma^{\mu}\,\nu^0_L\,-\,\overline{e}_L\,\gamma^{\mu}\,e^0_L
\,-2\sinˆ^{2}\theta_{W}\,J^{\mu}_{\text{e.m.}}\,\right]\,Z_{\mu}\,,
\end{equation}
are also invariant under the transformations given in eq.~\eqref{eq:diagU}.
\begin{equation}
\label{eq:nc}
\Lag'_{\text{\sc NC}}\,=\,\frac{g}{\cos\theta_W}\,\left[\,\overline{u}_L\,\gamma^{\mu}\,u_L\,-\,\overline{d}_L\,\gamma^{\mu}\,d_L
\,+\,
\,\overline{\nu}_L\,\gamma^{\mu}\,\nu_L\,-\,\overline{e}_L\,\gamma^{\mu}\,e_L
\,-2\sinˆ^{2}\theta_{W}\,J^{\mu}_{\text{e.m.}}\,\right]\,Z_{\mu}\,.
\end{equation}
Flavour changing neutral currents (FCNC) are naturally absent at
three-level in the SM, due to the GIM mechanism. Indeed ``charm'' was
invented in order to achieve this cancellation of FCNC.

\begin{exercise}
Suppose that ``charm'' did not exist, so that one would have 
\begin{equation}
\begin{pmatrix}u_1^0\\d_1^0\end{pmatrix}_L\,,\quad
{d_2}_L\,,\quad {u_1}_R\,,\quad {d_1}_R\,,\quad \quad {d_1}_R\,.
\end{equation}
Show that in this model FCNC automatically arise.
\end{exercise}

Historical note: Prior to the appearance of renormalisable gauge
interactions, physicists considered the possibility that weak neutral
currents could exist. However there was a \emph{strong prejudice}
against neutral currents due to the stringent experimental limits on the strength
of FCNC.

\begin{example}
  The decay $K^0_L\,\rightarrow\,\mu^{+}\,\mu^{-}$ has a branching
  ratio \emph{extremely suppressed}, with respect to the decay
  $K^0_L\,\rightarrow\,\pi^{+}\,e^{-}\,\overline{\nu_e}\,$. If FCNC
  existed they would have branching ratios of the same order of
  magnitude which are shown in figure~\ref{fig:KL}.

\begin{figure}
    \begin{center}
      \subfigure[fnm1][$K^{0}_L\longrightarrow\pi^{+}e^{-}\nu_e$]{  
       \begin{tikzpicture}[thick, 	level/.style={level distance=1.cm, line width=0.7mm}, node distance=0.8cm and 1.2cm]
       \coordinate[label=left:$d$] (s);
       \coordinate[above=of s,label=left:$s$] (d);
       \coordinate[right=of d](d1);
       \coordinate[right=of d1](d2);
       \coordinate[label=right:$u$,right=of d2](d3);
       \coordinate[label=right:$d$,below=of d3] (s1);  
       \coordinate[above=of d2](dw);
       \coordinate[above=of d3,label=right:$\overline{\nu}_e$](nue);
        \coordinate[above=of nue,label=right:$e$](e);
        \draw[fermion] (d) -- (d1);
         \draw[fermion] (d1) -- (d3);
        \draw[fermion] (s1) -- (s);
         \draw[gauge] (d1) -- node[label=above:$W^{-}$] {} (dw);
         \draw[fermion] (nue) -- (dw);
         \draw[fermion] (dw) -- (e);
	\end{tikzpicture}  
             }
      \qquad\qquad
      \subfigure[fnm2][Does not exist at tree-level in SM]{ 
        \begin{tikzpicture}[thick, 	level/.style={level distance=1.8cm, line width=0.7mm}, node distance=1.cm and 1.2cm]
          \coordinate[label=left:$d$] (s);
          \coordinate[below=of s] (sd);
          \coordinate[below=of sd,label=left:$s$] (d);
          \coordinate[right=of sd] (Wsd);
          \coordinate[right=of Wsd] (Wmm);
          \coordinate[right=of Wmm] (m);
          \coordinate[above=of m,label=right:$\mu^{-}$] (mm);
          \coordinate[below=of m,label=right:$\mu^{+}$] (mp);
          \draw[fermion] (Wsd) -- (d);
          \draw[fermion] (s) -- (Wsd);
          \draw[fermion] (mp) -- (Wmm);
          \draw[fermion] (Wmm) -- (mm); 
          \draw[gauge]   (Wsd) -- node[label=above:$Z$] {} (Wmm);
        \end{tikzpicture}  
      }
    \end{center}
    \caption{\label{fig:KL}}
  \end{figure}

\end{example}

From eq.~\eqref{eq:nc} we see that neutral current interactions
violate parity, since both couplings involving
$\overline{\psi}\gamma_{\mu}\psi$ and
$\overline{\psi}\gamma_{\mu}\gamma_{5}\psi$ are present.  

As a result of the GIM mechanism there are no tree-level contributions
to $K^0-\overline{K^{0}}$, $B^0-\overline{B^{0}}$,
$B_{S}-\overline{B_{S}}$ and $D^0-\overline{D^{0}}$ mixings. However
in the SM there are higher order contributions to these processes which are calculable. The
contributions from the diagrams given in figure~\eqref{fig:box} led to
the correct estimate to the charm quark mass~\cite{Gaillard:1974hs}
and the size of $B_{d}-\overline{B_{d}}$ mixing provided the first indirect evidence of a large top mass.

\begin{exercise}
Consider a simple extension of the SM which consists of the addition of an isosinglet quark $D$,
\begin{equation}
D_{L},D_{R}\,\sim\,(3,1,-1/3)\,.
\end{equation}
\renewcommand{\labelenumi}{\alph{enumi})}
\begin{enumerate}
\item Write down the most general quark mass terms which are obtained in the framework of this model.
\item Derive the structure of the charged currents.
\item Derive the structure of neutral currents, showing that there are FCNC in this model.
\item Show that although non-vanishing at tree level, FCNC are
  naturally suppressed in this model, provided the isosinglet quark
  $D$ is much heavier than the standard quarks.
\end{enumerate}
\end{exercise}

Neutral currents have played a crucial r\^ole in the construction of
the SM and its experimental tests and the discovery of Neutral weak
currents was the first great success of the SM. As it was here
described, the important feature of FCNC is that they are forbidden at
tree-level, both in the SM and in most of its extensions. At loop level FCNC are generated and have played a crucial r\^ole in
testing the SM and in putting bounds on New Physics beyond the SM through the study of process like:
$K^0-\overline{K^{0}}$, $B^0-\overline{B^{0}}$,
$B_{S}-\overline{B_{S}}$ and $D^0-\overline{D^{0}}$; rare kaon decays;
rare b-meson decays; CP violation. In this framework, SM contributes
to these processes at loop level and therefore New Physics has a
chance to give significant contributions.  On the other hand, the need
to suppress FCNC has lead to two dogmas: 
\begin{itemize}
\item[]\textbf{no $Z$-mediated FCNC at tree
level and no FCNC in the scalar sector, at tree level. }
\end{itemize}

S.~Glashow, S.~Weimberg~\cite{Glashow:1976nt} and
E.A.~Paschos~\cite{Paschos:1976ay} derived necessary and sufficient
conditions for having diagonal neutral currents, namely:
\renewcommand{\labelenumi}{\roman{enumi})}
\begin{enumerate}
\item All quarks of fixed charge and helicity must transform according
  to the same irreducible representation of $\mathsf{SU}(2)$ and
  correspond to the same eigenvalue of $T_3$.
\item All quarks should receive their contributions to the quark mass
  matrix from a single neutral scalar VEV.
\end{enumerate}

\textbf{Can one violate the above two dogmas in reasonable extensions of the SM? The
answer is yes! }

``Reasonable'' means that FCNC should be naturally suppressed without
fine-tuning. In the gauge sector, the \emph{dogma} can be violated
through the introduction of a $Q=1/3$ and/or $Q=2/3$ vector-like
quark~\cite{Branco:1986my,delAguila:1985ne,Branco:1992wr,Branco:1995us,Morozumi:1997af,Barenboim:1997qx,Barenboim:1997pf},
since in this model one has naturally small violation of $3\times3$
unitarity of the CKM matrix $V$ which in turn leads to $Z$-mediated
FCNC at tree level, which are naturally suppressed.

\begin{figure}[h]
  \begin{center}
    \subfigure[fnm1][$K^{0}-\overline{K^{0}}$ mixing]{  
      \begin{tikzpicture}[thick, node distance=1.7cm and 1.7cm]     
        \coordinate[label=left:$d$] (d);
        \coordinate[right=of d,] (dW);
        \coordinate[right=of dW] (Wd);
        \coordinate[below= of d,label=left:$s$] (s);
        \coordinate[right=of s] (sW);
        \coordinate[right=of sW] (Ws);
        \coordinate[right=of Wd,label=right:$d$] (d1);
        \coordinate[right=of Ws,label=right:$s$] (s1);
        \draw[fermion] (d) -- (dW);
        \draw[fermion] (sW) -- (s);
        \draw[fermion] (Ws) -- (s1);
        \draw[fermion] (d1) -- (Wd);
        \draw[gauge] (dW) -- node[label=above:$W$] {} (Wd);
        \draw[gauge] (sW) -- node[label=above:$W$] {} (Ws);
        \draw[fermion] (dW) -- node[label=left:{$u,c,t$}] {} (sW);
        \draw[fermion] (Wd) -- node[label=right:{$u,c,t$}] {}  (Ws);
      \end{tikzpicture}    }
    \qquad
    \subfigure[fnm2][$B^{0}_{d}-\overline{B^{0}_{d}}$ mixing]{ 
      \begin{tikzpicture}[thick, node distance=1.7cm and 1.7cm]
        \coordinate[label=left:$d$] (d);
        \coordinate[right=of d] (dW);
        \coordinate[right=of dW] (Wd);
        \coordinate[below= of d,label=left:$b$] (b);
        \coordinate[right=of b] (bW);
        \coordinate[right=of bW] (Wb);
        \coordinate[right=of Wd,label=right:$d$] (d1);
        \coordinate[right=of Wb,label=right:$b$] (b1);
        \draw[fermion] (d) -- (dW);
        \draw[fermion] (bW) -- (b);
        \draw[fermion] (d1) -- (Wd);
        \draw[fermion] (Wb) -- (Wb);
        \draw[fermion] (Wb) -- (b1);
        \draw[fermion] (dW) -- node[label=left:{$u,c,t$}] {} (bW);
        \draw[fermion] (Wd) -- node[label=right:{$u,c,t$}] {} (Wb);
        \draw[gauge]   (dW) -- node[label=above:$W$] {} (Wd);
        \draw[gauge]   (bW) -- node[label=above:$W$] {} (Wb);
      \end{tikzpicture}  
    }
  \end{center}
  \caption{\label{fig:box}}
\end{figure}

In the Higgs sector, the dogma can be violated and yet having FCNC automatically suppressed by small CKM matrix elements~\cite{Branco:1986my}.

\subsection*{Fundamental properties of the CKM matrix}

We have introduced in eq.~\eqref{eq:CKM} the CKM matrix~$V$, which
characterises the flavour changing charged currents in the quark
sector:
\begin{equation}
\Lag_{\text{CC}}\,=\,
\begin{pmatrix}
\overline{u} & \overline{c} & \overline{t}
\end{pmatrix}_L
\,\gamma^{\mu}\,
\begin{pmatrix}
V_{ud} & V_{us} & V_{ub}  \\
V_{cd} & V_{cs}   & V_{cb}     \\
V_{td} & V_{ts}   &  V_{tb}     
\end{pmatrix}
\,
\begin{pmatrix}
d \\ s \\ b
\end{pmatrix}_L
\,W_{\mu}^{+}\,
\,+\,\text{H.c.}
\,,
\end{equation}
The CKM matrix is complex, but some of its phases have no physical
meaning. This is due to the fact that one has the freedom to rephase
the  mass eigenstate quark fields $u_{\alpha}\,,d_{k}$:
\begin{equation}
u_{\alpha}\,=\,e^{i\,\varphi_{\alpha}}u'_{\alpha}\,,\qquad d_{k}\,=\,e^{i\,\varphi_{k}} d'_{k}\,.
\end{equation}
Under this rephasing one has:
\begin{equation}
\label{eq:rph}
V_{\alpha k}'\,=\,e^{i\,(\varphi_k-\varphi_{\alpha})}\,V_{\alpha k}\,.
\end{equation}
It is clear from eq.~\eqref{eq:rph} that the individual phases of $V_{ij}$ have no Physical meaning.
It is useful to look for rephasing invariant quantities, which do not change under this rephasing. The simplest examples are moduli $|V_{\alpha k}|$ and quartets $Q_{\alpha i \beta j}$, defined as
\begin{equation}
Q_{\alpha i \beta j}\,\equiv\,V_{\alpha i}\,V_{\beta j}\,V^*_{\alpha j}\,V^*_{\beta i}\,,
\end{equation}
with $\alpha\neq\beta$ and $i\neq j$. Invariants of higher
order may in general be written as functions of the quartets and the
moduli.

\begin{exercise}
Show that:
\begin{equation}
V_{\alpha i}\,V_{\beta j}\,V_{\gamma k}\,V^*_{\alpha j}\,V^*_{\beta k}\,V^*_{\gamma i}\,=\,\frac{Q_{\alpha i \beta j}\,Q_{\beta i \alpha   j}}{|V_{\beta i}|^2}\,.
\end{equation}
\end{exercise}

The quartets are easily constructed through the following scheme,
\begin{equation}
V\,=\,\begin{tikzpicture}[baseline=(U.center)]
\matrix [matrix of math nodes,left delimiter=(,right delimiter={)},nodes={outer sep=1pt}] (U) { 
V_{ud} &[7mm] \node (A) {V_{us}};   &[7mm]  \node (B) {V_{ub}};     \\[7mm]
\node (E) {V_{cd}}; &[7mm] \node (C) {V_{cs}};   &[7mm]  \node (D) {V_{cb}};     \\[7mm]
\node (F) {V_{td}}; &[7mm] \node (G) {V_{ts}};   &[7mm]  V_{tb}     \\
};
\draw [dotted, thick] (A) -- (D);
\draw [dotted, thick] (B) -- (C);
\draw [dotted, thick] (E) -- (G);
\draw [dotted, thick] (C) -- (F);
\end{tikzpicture}\,,
\end{equation}
where the two quartets, 
\begin{equation}
V_{us}\,V_{cb}\,V^*_{ub}\,V^*_{cs}\,=\,Q_{uscb}\,,\qquad
V_{cd}\,V_{ts}\,V^*_{td}\,V^*_{cs}\,=\,Q_{cdts}\,,
\end{equation}
are illustrated. The diagonal dotted line
refers to the product of the corresponding CKM elements.

\subsection{Neutrino masses}
In the SM, neutrinos are exactly massless.  No Dirac mass terms can be
written since right-handed neutrino fields are not introduced in the
SM. On the other hand, Majorana mass terms are not generated in higher
orders, due to exact $(B-L)$ conservation in the SM. As a result of
having massless neutrinos, neither leptonic mixing nor leptonic CP
violation can be generated in the SM. Indeed, any mixing arising from
the diagonalisation of the charged-lepton masses can be rotated away
by a redefinition of the neutrino fields.

In the view of above, one concludes that the discovery of leptonic
mixing and non-vanishing neutrino masses, rules out the SM, as it was
proposed. However a simple extension of the SM, sometimes denoted
$\nu$SM, can easily accommodate leptonic mixing and provide an
explanation for the smallness of neutrino masses, through the seesaw
mechanism~\cite{Minkowski:1977sc,Yanagida:1979as,GellMann:1980vs,Glashow:1980,Mohapatra:1979ia}. The
nature of neutrinos (i.e. Majorana or Dirac) is still an important
open question. Both in the case of
Majorana~\cite{Minkowski:1977sc,Yanagida:1979as,GellMann:1980vs,Glashow:1980,Mohapatra:1979ia}
or Dirac neutrinos~\cite{Branco:1978bz} one has to have a mechanism to
understand the smallness of neutrino masses.

\subsection{The Flavour sector of the SM} 

Let us now discuss the flavour sector of the SM. The gauge invariance
does not constrain the flavour structure of the Yukawa matrices $Y_u$,
$Y_d$ and using eq.~\eqref{eq:mdeff} one obtains two arbitrary mass
matrices $m_u$ and $m_d$.  The two quark mass matrices are arbitrary
complex matrices which need not to be
Hermitian~\cite{Branco:1994jx}. The two matrices $m_u$, $m_d$ contain
$(18+18)$ parameters, but most of them are not physical. Due to the
fermion family replication the gauge interaction part of
$\Lag_{\text{SM}}$ has a very large flavour symmetry. One can make
Weak-basis transformations which change $m_u$, $m_d$ but do not change
the physical content of $m_u$, $m_d$. One has then a large redundancy
in $m_u$, $m_d$. By making a WB transformation such as:
\begin{subequations}
\label{eq:WBT0}
\begin{align}
u^0_L\,&=\,W_L\,{u^0_L}'\,;\qquad u^0_R\,=\,W_R^u\,{u^0_R}'\,,\\
d^0_L\,&=\,W_L\,{d^0_L}'\,;\qquad d^0_R\,=\,W_R^d\,{d^0_R}'\,,
\end{align}
\end{subequations}
the gauge currents remain flavour diagonal but $m_u$, $m_d$ change as follows:
\begin{subequations}
\label{eq:WBTm}
\begin{align}
m_u\,\longrightarrow {m_u}'&=\,W_L^{\dagger}\,m_u\,W^u_R\,,\\
m_d\,\longrightarrow {m_d}'&=\,W_L^{\dagger}\,m_d\,W^d_R\,,
\end{align}
\end{subequations}
but the physical content does not change! Therefore, without loss of
generality, one can make a WB transformation so that $m_u$ is
diagonal, i.e. $m_u\,\diag(m_u,\,m_c,\,m_t)$ and $m_d$ is Hermitian
\begin{equation}
\label{eq:muH}
m_d\,=\,\begin{pmatrix}
m_{11} & m_{12} & m_{13} \\
m_{12}^{\ast} & m_{22} & m_{23} \\
m_{13}^{\ast} & m_{23}^{\ast} & m_{33}  
\end{pmatrix}\,.
\end{equation}
In this basis, the only rephasing invariant phase is
\begin{equation}
\varphi\,\equiv\, \arg(m_{12}\,m_{23}\,m_{13}^{\ast})\,,
\end{equation}
and there are ten independent parameters: 3 up-quark masses
$m_u,\,m_c,\,m_t$, 6 moduli down-type matrix elements $|{m_d}_{ij}|$
and one rephasing invariant phase $\varphi$.  There is to a difficulty in following a
bottom-up approach in the search for a solution to the \textbf{Flavour
  Puzzle}: even if there is a Flavour Symmetry behind the spectrum or
fermion masses and mixings, in what Weak-Basis will the symmetry be
transparent? For example Texture Zeroes are Weak-Basis dependent~\cite{Branco:1999nb}.

\subsection{CP Violation}

In order to study the CP properties of a Lagrangian, it is
convenient to separate the Lagrangian in two parts:
\begin{equation}
\label{eq:LagCP}
\Lag\,=\,\Lag_{\cp}\,+\,\Lag'\,,
\end{equation}
where $\Lag_{\cp}$ denotes the part of the Lagrangian which one knows
that conserves CP. At this stage it is important to recall that a pure
gauge Lagrangian is necessarily CP invariant~\cite{Grimus:1995zi}.
One should allow for \emph{the most general CP transformations}
allowed by $\Lag_{\text{CP}}$.  Typically, $\Lag_{\text{CP}}$ leaves a
large freedom of choice in the definition of CP transformations. CP is
violated if and only if there is no possible choice of CP
transformation which leaves the Lagrangian
invariant~\cite{Branco:1999fs}.  CP can be investigated in the fermion
mass eigenstate or in a weak basis.  We shall consider both cases.
Let us study the CP properties of the SM, after spontaneous gauge
symmetry breaking, and after diagonalisation of the quark mass
matrices, i.e.,
\begin{subequations}
\begin{align}
m_u\,&=\,\diag\left(m_u,m_c,m_t\right)\,,\\
m_d\,&=\,\diag\left(m_d,m_s,m_b\right)\,,
\end{align}
\end{subequations}
which are non-degenerate. In the mass eigenstate basis, the most general CP transformation is:
\begin{equation}
\label{eq:relCP}
\begin{aligned}
\cp\,{W^{+}}^{\mu}(t,\vec{r})\,\invcp\,&=\,-e^{i\zeta_W}{W^{-}}^{\mu}(t,-\vec{r})\,,\\
\cp\,{W^{-}}^{\mu}(t,\vec{r})\,\invcp\,&=\,-e^{-i\zeta_W}{W^{+}}^{\mu}(t,-\vec{r})\,,\\
\cp\,u_{\alpha}(t,\vec{r})\,\invcp\,&=\,e^{i\zeta_{\alpha}}\gamma^0\,C\,\overline{u}_{\alpha}^{\mathsf{T}}(t,-\vec{r})\,,\\
\cp\,d_{k}(t,\vec{r})\,\invcp\,&=\,e^{i\zeta_{k}}\gamma^0\,C\,\overline{u}_{k}^{\mathsf{T}}(t,-\vec{r})\,,
\end{aligned}
\end{equation}
where the conjugation matrix $C$ obeys to the relation
$\gamma_{\mu}C=-C\gamma^T_{\mu}$.  Invariance of charged current weak
interactions under CP constrains $V_{\alpha k}$ to satisfy the
following condition:
\begin{equation}
V_{\alpha k}^*\,=\,e^{i(\zeta_W+\zeta_{k}-\zeta_{\alpha})}\,V_{\alpha k}\,.
\end{equation}
If one considers a single element of the $CKM$ matrix $V$, the previous
condition can always be satisfied by using the freedom to choose
$\zeta_W,\zeta_{k},\zeta_{\alpha}\,$. However, it can be readily shown
that the condition constrains all quartets and all rephasing invariant
functions of $V$ to real. Therefore there is CP violation in the SM if
and only if any of the rephasing invariant functions of the CKM matrix
$V$ is not real. 

In the SM with $n_g$ generations, the CKM matrix $V$ is a $n_g\times
n_g$ unitary matrix and it can be then parametrised by $n_g^2$
independent parameters.  Through rephasing of quark fields, one can
remove $2\,n_g-1$ phases. Thus, the total number of parameters, denoted $N$, is
given by
\begin{equation}
N\,=\,n_g^2-2\,n_g-1\,=\,(n_g-1)^2\,,
\end{equation}
which shows that for three generations ($n_g=3$) one is left with 4 real
parameters. If one takes into account that a unitary matrix is
describe by $n_g(n_g-1)/2$ ``angles'', one can further count the total number of physical phases $N_{ph}$ as:
\begin{equation}
N_{ph}\,=\,N-\frac12n_g(n_g-1)\,=\,\frac12(n_g-1)(n_g-2)\,.
\end{equation}
We conclude that for 2 generations ($n_g=2$), there are no physical
phases left and therefore CP is conserved. In the case of three
generations ($n_g=3$), one has only one CP violating phase.  There is
another way of confirming this. For two generations, there is only one
rephasing invariant quartet $Q_{udcs}$, defined as
\begin{equation}
Q_{udcs}\,\equiv\,V_{ud}\,V_{cs}\,V_{us}^*\,V_{cd}^*\,.
\end{equation}
However using the orthogonality relation:
\begin{equation}
V_{ud}\,V_{cd}^*\,+\,V_{us}\,V_{cs}^*\,=\,0\,, 
\end{equation}
and multiplying by $V_{us}^{\ast}\,V_{cs}$, one obtains:
\begin{equation}
Q_{udcs}\,=\,-|V_{us}|^2\,|V_{cs}|^2\,,
\end{equation}
which shows that $Q_{udcs}$ is real.

\begin{figure}
\begin{center}
\begin{tikzpicture}[thick]
\coordinate (O) at (0,0);
\coordinate (A) at (2,3);
\coordinate (B) at (10,0);
\coordinate (BB) at (2,0);

\draw (O) -- (A) -- (B) -- (C) -- cycle;
\draw[-latex] (O) -- (A);
\draw[-latex] (A) -- (B);
\draw[-latex] (B) -- (O);
\draw [black, dashed] (BB) --(A);

\tkzLabelSegment[right=3pt](BB,A){$h$}
\tkzLabelSegment[above left=3pt](O,A){$V_{ud}\,V_{ub}^{\ast}$}
\tkzLabelSegment[below=3pt ](O,B){$V_{cd}\,V_{cb}^{\ast}$}
\tkzLabelSegment[above right=2pt](A,B){$V_{td}\,V_{tb}^{\ast}$}

\tkzLabelAngle[pos=1.4](O,B,A){$\beta$}
\tkzLabelAngle[pos=0.4](O,A,B){$\alpha$}
\tkzLabelAngle[pos=0.6](A,O,B){$\gamma$}
\end{tikzpicture}
\end{center}
\caption{\label{fig:tri} Unitarity triangle}
\end{figure}
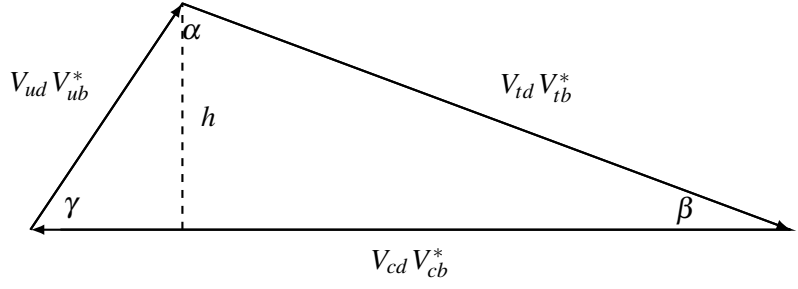

Considering now the case of three generations, we see that  orthogonality of the first two rows of $V$ leads to
\begin{equation}
V_{ud}\,V_{cd}^*\,+\,V_{us}\,V_{cs}^*\,+\,V_{ub}\,V_{cb}^*\,=\,0\,.
\end{equation}
Multiplying by $V_{us}^*\,V_{cs}$ and taking imaginary parts one obtains:
\begin{equation}
\imag Q_{udcs}\,=\, -\imag Q_{ubcs}\,.
\end{equation}
In an analogous way, one can show that for $n_g=3$ the imaginary parts
of all quartets are equal, up to a sign. In the SM with
three generations $|\imag Q|$ gives the strength of CP violation. If
we consider the orthogonality between the first and third columns of
$V$:
\begin{equation}
V_{ud}\,V_{ub}^*\,+\,V_{cd}\,V_{cb}^*\,+\,V_{td}\,V_{tb}^*\,=\,0\,.
\end{equation}
This equation may be interpreted as a "triangle" as represented in
figure~\ref{fig:tri}.  One verifies easily that under rephasing, the
triangle rotates. Therefore the orientation of the triangle has no
physical meaning. Obviously, the internal angles of the triangles are
rephasing invariant, namely
\begin{subequations}
\begin{align}
\alpha & \equiv\,\arg\left[-V_{td}\,V_{ub}\,V_{ud}^{\ast}\,V_{tb}^{\ast}\right]\,=\,
\arg(-Q_{ubtd})\,,\\
\beta & \equiv\,\arg\left[-V_{cd}\,V_{tb}\,V_{cb}^{\ast}\,V_{td}^{\ast}\right]\,=\,
\arg(-Q_{tbcd})\,,\\
\label{eq:gdef}
\gamma & \equiv\,\arg\left[-V_{ud}\,V_{cb}\,V_{ub}^{\ast}\,V_{cd}^{\ast}\right]\,=\,
\arg(-Q_{cbud})\,,
\end{align}
\end{subequations} 
and one gets the following relation
\begin{equation}
\alpha\,+\,\beta\,+\,\gamma\,=\,\arg(-1)\,=\,\pi\pmod{\pi}\,.
\end{equation}
This is true "by definition", and therefore it is not a test of unitarity!!

The quantity $\imag Q$ has a simple geometrical interpretation. It is
twice the area of the unitarity triangles, as sketched in
figure~\ref{fig:tri}. The area of the triangles, $A$, is given by
\begin{equation}
A\,=\,\left|V_{cd}\,V_{cb}^{\ast}\right|\,\frac{h}2\,,
\end{equation}
where the height of triangle, $h$, is given by
\begin{equation}
h\,=\,\left|V_{ud}\,V_{ub}^{\ast}\right|\,\sin\gamma\,,
\end{equation}
with $\gamma$ defined in eq.~\eqref{eq:gdef}. One then obtains
\begin{equation}
A\,=\,\frac12\left|\imag Q_{udcb}\right|\,.
\end{equation}
Since all $|\imag Q|$ are equal then all triangles have the same area.

Experimentally we know that:
\begin{equation}
\left|V_{\rm{CKM}}\right|\,\simeq\,\begin{pmatrix}
1 & \lambda & \lambda^3\\
 \lambda & 1 &  \lambda^2\\
  \lambda^3 &  \lambda^2 & 1
  \end{pmatrix}\,,
\end{equation}
with $\lambda \approx 0.22$.

The six unitarity triangles are given by
\begin{equation}
\begin{aligned}
\stackrel{\lambda}{V_{ud}\,V_{us}^{\ast}} \,+\,
\stackrel{\lambda}{V_{cd}\,V_{cs}^{\ast}}\,+\,
\stackrel{\lambda^5}{V_{td}\,V_{ts}^{\ast}} &=\,0\,,\qquad T_{ds}
\\
\stackrel{\lambda^3}{V_{ud}\,V_{ub}^{\ast}} \,+\,
\stackrel{\lambda^3}{V_{cd}\,V_{cb}^{\ast}}\,+\,
\stackrel{\lambda^3}{V_{td}\,V_{tb}^{\ast}} &=\,0\,,\qquad T_{db}
\\
\stackrel{\lambda^4}{V_{us}\,V_{ub}^{\ast}} \,+\,
\stackrel{\lambda^2}{V_{cs}\,V_{cb}^{\ast}}\,+\,
\stackrel{\lambda^2}{V_{ts}\,V_{tb}^{\ast}} &=\,0\,,\qquad T_{sb}
\\
\stackrel{\lambda}{V_{ud}\,V_{cd}^{\ast}} \,+\,
\stackrel{\lambda}{V_{cs}\,V_{cs}^{\ast}}\,+\,
\stackrel{\lambda^5}{V_{ub}\,V_{cb}^{\ast}} &=\,0\,,\qquad T_{uc}
\\
\stackrel{\lambda^3}{V_{ud}\,V_{td}^{\ast}} \,+\,
\stackrel{\lambda^3}{V_{us}\,V_{ts}^{\ast}}\,+\,
\stackrel{\lambda^3}{V_{ub}\,V_{td}^{\ast}} &=\,0\,,\qquad T_{ut}
\\
\stackrel{\lambda^4}{V_{cd}\,V_{td}^{\ast}} \,+\,
\stackrel{\lambda^2}{V_{cs}\,V_{ts}^{\ast}}\,+\,
\stackrel{\lambda^2}{V_{tb}\,V_{tb}^{\ast}} &=\,0\,,\qquad T_{ct}
\end{aligned}
\end{equation}

Let us now comment on the strength of CP violation in the SM, which 
\begin{equation}
\left|\imag Q\right|\,=\,\left|\,
\stackrel{\lambda^0}{V_{ud}}\,
\stackrel{\lambda^3}{V_{ub}}\,
\stackrel{\lambda}{V_{cd}}\,
\stackrel{\lambda^2}{V_{cb}}\,
\right|\,\sin\gamma\,.
\end{equation}
In order to account for CP violation in the kaon sector, $\sin\gamma$ should be of order 1. So $|\imag Q|\approx\lambda^6$.

The strength of CP violation (measured by $\imag Q$) is small in the SM, due to the smallness of some CKM moduli $|V_{ij}|$, like $|V_{ub}|$, $|V_{cb}|$. What would be the maximal possible value of  $\imag Q$? The maximal value is obtained for the following mixing matrix  with universal moduli as
\begin{equation}
V\,=\,\frac1{\sqrt3}\,\begin{pmatrix}
1 & 1 & 1\\
1 & \omega & \omega^{\ast}\\
1 & \omega^{\ast} & \omega
\end{pmatrix}\,,
\end{equation} 
 with $\omega\equiv\exp(i 2\pi/3)\,$, yielding
 \begin{equation}
 \imag Q\,=\,\frac1{6\sqrt3}\,\approx\,0.096\,. 
 \end{equation}
 
 A convenient parametrisation of the CKM matrix is the so-called
 Standard Parametrisation, which is defined by the product of three
 rotations, namely:
 \begin{equation}
 \begin{split}
 V(\theta_{12},\theta_{13},\theta_{23},\delta_{13})&\,=\,
 \begin{pmatrix}
 1 & 0 & 0\\
 0 & c_{23} & s_{23}\\
 0 & -s_{23} & c_{23}
 \end{pmatrix}
 \begin{pmatrix}
 c_{13} & 0 & s_{13}\,e^{-i\delta_{13}}\\
 0 & 1 & 0\\
 -s_{13}\,e^{i\delta_{13}} & 0 & c_{13} 
 \end{pmatrix}
 \begin{pmatrix}
 c_{12} & s_{12} & 0 \\
 -s_{12} & c_{12} & 0\\
 0 & 0 & 1
 \end{pmatrix}
 \\ \,=\,&
 \begin{pmatrix}
c_{12} \,c_{13} & s_{12}\,c_{13} & s_{13}\,e^{-i\delta_{13}} \\
 -s_{12}\,c_{23}-c_{12}\,s_{23}\,s_{13}e^{i\delta_{13}}   & 
  c_{12}\,c_{23}-s_{12}\,s_{23}\,s_{13}e^{i\delta_{13}}   & s_{23} \,c_{13}\\
  s_{12}\,s_{23}-c_{12}\,c_{23}\,s_{13}e^{i\delta_{13}}  &
 -c_{12}\,s_{23}-s_{12}\,c_{23}\,s_{13}e^{i\delta_{13}}  & c_{23} \,c_{13}
 \end{pmatrix}\,,
 \end{split}
 \end{equation}
 where $c_{ij}\equiv\cos\theta_{ij}$ and
 $s_{ij}\equiv\sin\theta_{ij}$. One of the advantages of the Standard
 Parametrisation is that the $s_{ij}$ are simply related to directly
 measured quantities:
 \begin{equation}
 s_{13}\,=\,|V_{ub}|\,,\qquad
 s_{12}\,=\,\frac{|V_{us}|}{\sqrt{1-|V_{ub}|^2}}\,,\qquad
 s_{23}\,=\,\frac{|V_{cb}|}{\sqrt{1-|V_{ub}|^2}}\,.
 \end{equation}
 Once $s_{ij}$ are fixed, all data has to be fit by a single parameter: $\delta_{13}\,$.
 
\subsection*{Invariant Approach to CP Violation}

In this section we review the invariant approach to CP
violation~\cite{Bernabeu:1986fc}. As previously indicated we write
$\Lag$ as in eq.~\eqref{eq:LagCP}. In order to analyse whether the
whole Lagrangian violates CP, one has to check whether the CP
transformation under which $\Lag_{\cp}$ is invariant implies
non-trivial restrictions, i.e. restrictions which may not be satisfied
by $\Lag'\,$ defined in eq.~\eqref{eq:LagCP}.  In the case of the SM,
the most general CP transformations which leave $\Lag_{\cp}$ invariant
are:
\begin{equation}
\label{eq:relCPa}
\begin{aligned}
\cp\,u_L^0\,\invcp\,&=\,e^{i\zeta_W}\,K_L\,\gamma^0\,C\,\overline{u^0_L}^{\mathsf{T}}\,,\\
\cp\,d_L^0\,\invcp\,&=\,K_L\,\gamma^0\,C\,\overline{d^0_L}^{\mathsf{T}}\,,\\
\cp\,u_R^0\,\invcp\,&=\,K^u_R\,\gamma^0\,C\,\overline{u^0_R}^{\mathsf{T}}\,,\\
\cp\,d_R^0\,\invcp\,&=\,K^d_R\,\gamma^0\,C\,\overline{d^0_R}^{\mathsf{T}}\,,
\end{aligned}
\end{equation}
where $K_L$, $K^u_L$ and $K^d_R$ are unitary matrices acting in
flavour space. It can be shown that in order for the Yukawa
interactions (or equivalently $m_u\,,m_d$) to be CP invariant, the
following relations have to be satisfied:
\begin{equation}
\begin{aligned}
\label{eq:mm}
K^{\dagger}_L\,m_u\,K^u_R&=m^{\ast}_u\,,\\
K^{\dagger}_L\,m_d\,K^d_R&=m^{\ast}_d\,.
\end{aligned}
\end{equation}
The existence of the matrices  $K_L$, $K^u_L$ and $K^d_R$ 
is a necessary and sufficient condition for CP invariance in the SM.

\begin{exercise}
Prove the above result.
\end{exercise}

It is rather convenient to define the Hermitian matrices $H_u$,$H_d$ as:
\begin{equation}
H_{u}\,\equiv\,m_{u}\,m^{\dagger}_{u}\,,\qquad
H_{d}\,\equiv\,m_{d}\,m^{\dagger}_{d}\,.
\end{equation}
Thus, from eq.~\eqref{eq:mm} one derives:
\begin{equation}
\begin{aligned}
K^{\dagger}_L\,H_u\,K_L\,&=\, H^{\ast}_u\,=\,H^{\mathsf{T}}_u\,,\\ 
K^{\dagger}_L\,H_d\,K_L\,&=\, H^{\ast}_u\,=\,H^{\mathsf{T}}_u\,,
\end{aligned} 
\end{equation}
and therefore one has 
\begin{equation}
K^{\dagger}_L\,[H_u,H_d]\,K_L\,=\,[H_u^{\mathsf{T}},H_d^{\mathsf{T}}]\,=\,-[H_u,H_d]^{\mathsf{T}}\,.
\end{equation} 
Taking the trace of an odd $r$ power of the above equation we find
\begin{equation}
\label{CPINV}
\tr[H_u,H_d]^r\,=\,0\,.\qquad(r\;\,\rm{odd}) 
\end{equation}
Therefore, one concludes that in the SM CP invariance implies
$\tr[H_u,H_d]^r\,=\,0\,$. For the case of $r=1$ this relation is
trivially satisfied, since the trace of a commutator is automatically
zero. The minimum non-trivial case is for $r=3$. Note that
eq.~\eqref{CPINV} is a necessary condition for CP invariance for any
number of generations. For two generations the invariant automatically
vanishes. In the case of three generations one obtains:
\begin{equation}
\label{eq:I}
\begin{split}
\tr[H_u,H_d]^3=6i&(m^2_t-m^2_c)(m^2_t-m^2_u)(m^2_c-m^2_u)\\
&(m^2_b-m^2_s)(m^2_b-m^2_d)(m^2_s-m^2_d)\,\imag Q.
\end{split}
\end{equation}
It can be shown~\cite{Bernabeu:1986fc} that the vanishing of $\tr[H_u,H_d]^3$ is a necessary and sufficient condition for CP invariance in the SM with three generations. For more than three generations the vanishing of this invariant continuous being a necessary condition for CP invariance, but no longer is a sufficient condition.  In the  case of three generations one has\footnote{For any
$3\times3$ traceless matrix $A$: $\tr A^3=3\,\det(A)$.}: 
\begin{equation}
\det[H_u,H_d]\,=\,\frac13\, tr[H_u,H_d]^3\,,
\end{equation}
and thus the vanishing of the above determinant can be
used~\cite{Jarlskog:1985ht} as a necessary and sufficient condition
for CP invariance. .

\begin{exercise}
  Prove the above result given by eq.~\eqref{eq:I}. Hint: choose to
  work in a weak basis where either $H_u$ or $H_d$ is diagonal.
\end{exercise}

\section{Physics Beyond the Standard Model}

\subsection{Neutrino Masses}

It was already mentioned that in the SM, neutrinos are strictly
massless.  There are no Dirac mass terms since no fermionic
right-handed field $\nu_R$ is not introduced. There are no Majorana
mass terms at tree level, since no scalar $\mathsf{SU}(2)$-triplet is
introduced.  Also no Majorana mass is generated at higher orders, due
to exact $B-L$ conservation. Note that the Majorana mass term:
\begin{equation}
\nu_L^{\mathsf{T}}\,C\,\nu_L\,,
\end{equation}
violates $B-L$ by 2 units. The same applies to $\mathsf{SU}(5)$ GUT, where $B-L$ is also an accidental symmetry.

It can be readily seen that in the SM, there is no leptonic mixing. The leptonic charged currents is given by
\begin{equation}
\overline{\nu}^0_{Li}\,\gamma^{\mu}\,\ell^0_{Li}\,W_{\mu}\,,
\end{equation}
where $\nu^0_{Li}$ and $\ell^0_{Li}$ are weak eigenstates. After diagonalisation of the charged lepton mass matrix the leptonic charged currents become 
\begin{equation}
\overline{\nu}^0_{L}\,\gamma^{\mu}\,U\,\ell_{L}\,W_{\mu}\,,
\end{equation}
where $\ell_{L}$ is now in the mass eigenstate. But the unitary matrix
$U$ can be eliminated through a redefinition of $nu^0_{L}$ so that the
charged currents become flavour diagonal:
\begin{equation}
\overline{\nu}_{L}\,\gamma^{\mu}\,\ell_{L}\,W_{\mu}\,.
\end{equation}

Observation of neutrino oscillations provides clear evidence for New Physics beyond the SM. The Minimal extension of the SM which allows for non-vanishing neutrino masses introduces right-handed neutrino fields, 
$\nu_{Ri}$, a "strange" missing feature of the SM. Once the right-handed neutrino fields are introduced, Dirac masses for neutrinos are generated. Yukawa interactions can then be written as
\begin{equation}
{(Y_{\nu})}_{ij}\,\overline{\ell}_i\,\tilde{\phi}\,\nu_{Ri}\,+\,\text{H.c.}\,,
\end{equation} 
where the Yukawa mass matrix $Y_{\nu}$ is arbitrarily complex. After the spontaneous breaking of the electroweak  gauge symmetry  Dirac neutrino masses are then generated as
 \begin{equation}
m_D\,=\,{(Y_{\nu})}_{ij}\,\frac{v}{\sqrt2}\,.
\end{equation} 
If one writes the most general Lagrangian consistent with gauge invariance and renormalisability, one has to include the mass term:
\begin{equation}
{(M_{R})}_{ij}\,\nu_{Ri}^{\mathsf{T}}\,C\,\nu_{Rj}\,,
\end{equation}
where $M_{R}$ is a symmetric complex matrix. One may have $M_R\gg v$, since the mass term is gauge invariant. This leads to the seesaw mechanism, with:
\begin{subequations}
\begin{align}
(m_{\nu})_{\text{light}}\,&\approx\,\frac{v^2}{M_R}\,,\\
(M)_{\text{heavy}}\,&\approx\,M_R\,.
\end{align}
\end{subequations}
Note that this minimal extension of the SM, sometimes denoted $\nu$SM, is actually "simpler" and more "natural" than the SM, providing a simple and plausible explanation for the smallness of neutrino masses. 

For the moment let us consider the low energy limit of the $\nu$SM. Let us consider the neutrino masses and mixing at low energies, i.e the mass term Lagrangian
\begin{equation}
\Lag_{\text{mass}}\,=\,-\overline{\ell}_L\,m_{\ell}\,\ell_R\,-\,\frac12\nu_L^{\mathsf{T}}\,C\,m_{\nu}\,\nu_L\,+\,\text{H.c.}\,,
\end{equation}
and the charged current Lagrangian 
\begin{equation}
\Lag_{W}\,=\,\frac{g}{\sqrt2}\,\overline{\ell}_L\,\gamma^{\mu}\,\nu_L\,W_{\mu}\,+\,\text{H.c.}\,,
\end{equation}
where $m_{\ell}$, an arbitrary complex matrix, and $m_{\nu}$, a symmetric complex matrix, encode all information about lepton masses and mixing. There is a great redundancy in $m_{\ell}$, $m_{\nu}$, since not all their parameters are physical. This redundancy stems from the freedom to make Weak-Basis transformations:
\begin{equation}
\nu_L=W_L\,\nu_L´\,, \qquad \ell_L=W_L\,\ell_L´\,, \qquad \ell_R=W_R\,\ell_R´\,,
\end{equation}
where $W_L$, $W_L$ are unitary matrices. The matrices $m_{\ell}$, $m_{\nu}$ transform then as:
\begin{equation}
m^{\prime}_{\ell}\,=\,W^{\dagger}_L\,m_{\ell}\,W_R\,, \qquad  
m^{\prime}_{\nu}\,=\, W^{\mathsf{T}}_L\,m_{\nu}\,W_L\,.
\end{equation}

One can use the freedom to make WB transformations to go to a basis where 
\begin{equation}
m_{\ell}\,=\,d_{\ell}
\end{equation}
is diagonal and real.  In this basis, one can still make a rephasing: 
\begin{equation}
  \ell_{L,R}^{\prime\prime}=K_L\,\ell_{L,R}´\,, \qquad \nu^{\prime\prime}_L=K_L\,\nu_L´\,,
\end{equation}
with $K_L\,=\,\diag(e^{i\,\varphi_1},\,e^{i\,\varphi_2},\,e^{i\,\varphi_3})\,$. Under this rephasing $d_{\ell}$ remains invariant, but $m_{\nu}$ transforms as:
\begin{equation}
\left(m_{\nu}^{\prime\prime}\right)_{ij}\,=\, e^{i\,(\varphi_i\,+\,\varphi_j)}\,\left(m_{\nu}^{\prime}\right)_{ij}\,.
\end{equation}
One can eliminate $n$ phases from $m_{\nu}$. The number of physical phases in $m_{\nu}$ is:
\begin{equation}
N_{ph} \,=\, \frac12\,n(n+1)-n \,=\, \frac12\,n(n-1)\,,
\end{equation}
where $n$ is the number of right-handed neutrino fields. For $n=3$ one has $N_{ph}=3$. So altogether one has in $m_{\nu}$:  6 real moduli $|(m_{\nu}^{\prime\prime})_{ij}|$ and three phases ($N_{ph}=3$). The individual phases of $(m_{\nu}^{\prime\prime})_{ij}$ have no physical meaning because they are not rephasing invariant. But one can construct polynomials of $(m_{\nu}^{\prime\prime})_{ij}$ which are rephasing invariant. Examples of rephasing invariant polynomials:
\begin{equation}
\begin{aligned}
P_1 & \,\equiv\,(m_{\nu}^{\ast})_{11}(m_{\nu}^{\ast})_{22}(m_{\nu})^2_{12}\,,\\
P_2 & \,\equiv\,(m_{\nu}^{\ast})_{11}(m_{\nu}^{\ast})_{33}(m_{\nu})^2_{13}\,,\\
P_3 & \,\equiv\,(m_{\nu}^{\ast})_{33}(m_{\nu}^{\ast})_{12}(m_{\nu})_{13}(m_{\nu})_{23}\,.
\end{aligned}
\end{equation}
Let us now discuss the generation of the  leptonic mixing in the charged current. The charged leptonic mass matrix is diagonalised as
\begin{equation}
  U_{eL}^{\dagger}m_{\ell}\,U_{eR}\,=\,\diag(m_e\,,
  m_{\mu}\,, m_{\tau})\,,
\end{equation}
through the unitary matrices $U_{eL}\,$, $U_{eL}\,$, while the effective neutrino mass matrix is diagonalised as
\begin{equation}
 U_{\nu}^{\mathsf{T}}\,m_{\nu}\,U_{\nu}\,=\,\diag(m_1\,,m_2\,,m_3)\,.
\end{equation}
through the unitary matrix  $U_{\nu}\,$. Under these unitary transformations the charged currents are
\begin{equation}
\Lag_W\,=\,\frac{g}{\sqrt2}\,\overline{\ell}_L\,\gamma^{\mu}\,U\,
\nu_L\,W_{\mu}\,+\,\text{H.c.}\,.
\end{equation}
The unitary matrix $U$, which measure the mixing in the leptonic sector, given by
\begin{equation}
U\,\equiv\,U_{\ell}^{\dagger}\, U_{\nu}\,,
\end{equation}
is the so-called the Pontecorvo-Maki-Nakagawa-Sakata~(PMNS)~\cite{Pontecorvo:1957cp,Pontecorvo:1957qd,Maki:1962mu}
  matrix. In this basis, there is still freedom to rephase the charged lepton fields
 \begin{equation}
  \ell_j\,\longrightarrow\,\ell^{\prime}_j\,=\,\\exp(,i\,\phi_j),\ell_{j}\,,
\end{equation} 
with arbitrary phases $\phi_j$. Due to the Majorana nature of neutrinos, the rephasing:
\begin{equation}
\nu_{Lk}\,\longrightarrow\,\nu^{\prime}_{Lk}\\,=\,\\exp(,i\,\psi_j),\nu_{Lk}\,,
\end{equation}
with arbitrary phases $\psi_k$, is not allowed, since it would not leave the Majorana mass terms
\begin{equation}
\nu_{Lk}^{\mathsf{T}}\,C\,m_k\,\,\nu_{Lk}\,,
\end{equation}
invariant. In the mass eigenstate basis, the charred currents are
\begin{equation}
\Lag_W\,=\,\frac{g}{\sqrt2}\,
\begin{pmatrix}
\overline{e} & \overline{\mu} & \overline{\tau}
\end{pmatrix}_L
\,\gamma^{\mu}\,
\begin{pmatrix}
U_{e1} & U_{e2} & U_{e3}\\
U_{\mu1} & U_{\mu2} & U_{\mu3}\\
U_{\tau1} & U_{\tau2} & U_{\tau3}
\end{pmatrix}
\,
\begin{pmatrix}
\nu_1 \\ \nu_2 \\ \nu_3
\end{pmatrix}_L
\,W_{\mu}\,+\,\text{H.c.}\,.
\end{equation}
For the moment we do not introduce the constraints of $3\times3$ unitarity. Note that in the context of type-one seesaw the PMNS matrix is not unitary.

\subsubsection*{Rephasing invariant quantities}

Let us recall the situation in the quark sector where the charged current are given in the mass eigenstate basis as
\begin{equation}
\Lag_W\,=\,\frac{g}{\sqrt2}\,
\begin{pmatrix}
\overline{u} & \overline{c} & \overline{t}
\end{pmatrix}_L
\,\gamma^{\mu}\,
\begin{pmatrix}
V_{ud} & V_{us} & V_{ub}\\
V_{cd} & V_{cs} & V_{cb}\\
V_{td} & V_{ts} & V_{tb}
\end{pmatrix}
\,
\begin{pmatrix}
d \\ s \\ b
\end{pmatrix}_L
\,W_{\mu}\,+\,\text{H.c.}\,.
\end{equation} 
The CKM matrix $V$, has 9 moduli and the 4 rephasing invariant phases defined as
\begin{subequations}
\begin{align}
\beta&\,\equiv\,\arg\left(-V_{cd}\,V_{tb}\,V_{cb}^{\ast}\,V_{td}^{\ast}\right)\,,\\
\gamma&\,\equiv\,\arg\left(-V_{ud}\,V_{cb}\,V_{ub}^{\ast}\,V_{cd}^{\ast}\right)\,,\\
\chi\,=\,\beta_s&\,\equiv\,\arg\left(-V_{cb}\,V_{ts}\,V_{cs}^{\ast}\,V_{tb}^{\ast}\right)\,,\\
\chi'&\,\equiv\,\arg\left(-V_{us}\,V_{cd}\,V_{ud}^{\ast}\,V_{cs}^{\ast}\right)\,.
\end{align}
\end{subequations}

A novel feature in the leptonic sector with Majorana neutrinos, is the presence of rephasing invariant bilinear:
\begin{equation}
\arg\left(U_{\ell\alpha}\,U^{\ast}_{\ell\beta}\right)\,,
\end{equation}
where there is no summation of repeated indices. These are the so-called \textbf{Majorana-type phase}. There are six independent Majorana-type phase. This is true even when unitarity is not imposed on the PMNS matrix $U$.  It applies to a general framework with an arbitrary number of right-handed neutrinos. A possible choice for the six independent Majorana-type phases is~\cite{Branco:2008ai}:
\begin{subequations}
\label{eqs:Majorana}
\begin{align}
\beta_1&\,\equiv\,\arg\left(U_{e1},\,U^{\ast}_{e2}\right)\,,\qquad 
\gamma_1\,\equiv\,\arg\left(U_{e1},\,U^{\ast}_{e3}\right)\,,\\
\beta_2&\,\equiv\,\arg\left(U_{\mu1},\,U^{\ast}_{\mu2}\right)\,,\qquad 
\!\!\!\!\gamma_2\,\equiv\,\arg\left(U_{\mu1},\,U^{\ast}_{\mu3}\right)\,,\\
\beta_3&\,\equiv\,\arg\left(U_{\tau1},\,U^{\ast}_{\tau2}\right)\,,\qquad 
\gamma_3\,\equiv\,\arg\left(U_{\tau1},\,U^{\ast}_{\tau3}\right)\,.
\end{align}
\end{subequations}
One can choose the following four independent Dirac-type invariant phases:
\begin{subequations}
\begin{align}
\sigma^{12}_{e\mu} & \,\equiv\,
\arg\left(U_{e1}\,U_{\mu2}\,U^{\ast}_{e2}\,U^{\ast}_{\mu1}\right)\,=\,\beta_1-\beta_2\,,\\
\sigma^{12}_{e\tau} & \,\equiv\,
\arg\left(U_{e1}\,U_{\tau2}\,U^{\ast}_{e2}\,U^{\ast}_{\tau1}\right)\,=\,\beta_1-\beta_3\,,\\
\sigma^{13}_{e\mu} & \,\equiv\,
\arg\left(U_{e1}\,U_{\mu3}\,U^{\ast}_{e3}\,U^{\ast}_{\mu1}\right)\,=\,\gamma_1-\gamma_2\,,\\
\sigma^{13}_{e\tau} & \,\equiv\,
\arg\left(U_{e1}\,U_{\tau3}\,U^{\ast}_{e3}\,U^{\ast}_{\tau1}\right)\,=\,\gamma_1-\gamma_3\,.
\end{align}
\end{subequations}
If one assumes $3\times3$ unitarity of the PMNS matrix, the full leptonic mixing matrix can be reconstructed~\cite{Branco:1999nb,Branco:2008ai,Botella:2002fr} from the six independent Majorana phases given in eqs.~\eqref{eqs:Majorana}. Normalisation of rows  and columns plays an important r\^ole. It prevents the blowing up"of unitarity triangles. For three generations and assuming $3\times3$ unitarity of the PMNS matrix can be parametrised by:
\begin{equation}
U\,=\,V\,K\,,
\end{equation}
where $V$ is parametrised through the Standard Parametrisation:
\begin{equation}
\label{eq:spn}
V(\theta_{12},\theta_{13},\theta_{23},\delta_{13})\,=\,
 \begin{pmatrix}
c_{12} \,c_{13} & s_{12}\,c_{13} & s_{13}\,e^{-i\delta_{13}} \\
 -s_{12}\,c_{23}-c_{12}\,s_{23}\,s_{13}e^{i\delta_{13}}   & 
  c_{12}\,c_{23}-s_{12}\,s_{23}\,s_{13}e^{i\delta_{13}}   & s_{23} \,c_{13}\\
  s_{12}\,s_{23}-c_{12}\,c_{23}\,s_{13}e^{i\delta_{13}}  &
 -c_{12}\,s_{23}-s_{12}\,c_{23}\,s_{13}e^{i\delta_{13}}  & c_{23} \,c_{13}
 \end{pmatrix}\,,
 \end{equation}
 and the phase matrix $K$ given by
$K\,=\,\diag(1,\,e^{i\alpha_1/2},\,e^{i\alpha_2/2})\,$.
One can eliminate the phase $\delta$ from the first row by writing:
\begin{equation}
\label{eq:spn1}
U\,=\,V\,K'\,,
\end{equation}
where 
\begin{equation}
K'\,=\,\diag(1,\,1,\,e^{i\delta})\,K\,,
\end{equation}
convenient for the analysis of the neutrinoless double-beta decay ($0\nu\beta\beta$). The unitarity triangles in the leptonic sector, within the hypothesis of unitarity of the PMNS matrix, can be split into two category:
Dirac unitarity triangles
\begin{equation}
\begin{aligned}
U_{e1}\,U_{\mu 1}^{\ast}\,+\,
U_{e2}\,U_{\mu2}^{\ast}\,+\,
U_{e3}\,U_{\mu3}^{\ast} &=\,0\,,\qquad T_{e\mu}
\\
U_{e1}\,U_{\tau1}^{\ast}\,+\,
U_{e2}\,U_{\tau2}^{\ast}\,+\,
U_{e3}\,U_{\tau3}^{\ast} &=\,0\,,\qquad T_{e\tau}
\\
U_{\mu1}\,U_{\tau1}^{\ast}\,+\,
U_{\mu2}\,U_{\tau2}^{\ast}\,+\,
U_{\mu3}\,U_{\tau3}^{\ast} &=\,0\,,\qquad T_{\mu\tau}
\end{aligned}
\end{equation}
Majorana unitarity triangles
\begin{equation}
\begin{aligned}
U_{e1}\,U_{e2}^{\ast}\,+\,
U_{\mu1}\,U_{\mu2}^{\ast}\,+\,
U_{\tau1}\,U_{\tau2}^{\ast} &=\,0\,,\qquad T_{12}
\\
U_{e1}\,U_{e3}^{\ast}\,+\,
U_{\mu1}\,U_{\mu3}^{\ast}\,+\,
U_{\tau1}\,U_{\tau3}^{\ast} &=\,0\,,\qquad T_{13}
\\
U_{e2}\,U_{e3}^{\ast}\,+\,
U_{\mu2}\,U_{\mu3}^{\ast}\,+\,
U_{\tau2}\,U_{\tau3}^{\ast} &=\,0\,,\qquad T_{23}
\end{aligned}
\end{equation}

\begin{figure}
\begin{center}
\begin{tikzpicture}[thick]
\coordinate (O) at (0,0);
\coordinate (A) at (2,3);
\coordinate (B) at (10,0);

\draw (O) -- (A) -- (B) -- (C) -- cycle;
\draw[fermion] (O) -- (A);
\draw[fermion] (A) -- (B);
\draw[fermion] (B) -- (O);

\tkzLabelSegment[above left=3pt](O,A){$U_{e1}\,U_{e3}^{\ast}$}
\tkzLabelSegment[below=3pt ](O,B){$U_{\mu1}\,U_{\mu3}^{\ast}$}
\tkzLabelSegment[above right=2pt](A,B){$U_{\tau1}\,U_{\tau3}^{\ast}$}

\tkzLabelAngle[pos=0.6](A,O,B){$\varphi$}
\end{tikzpicture}
\end{center}
\caption{\label{fig:Majtri} Unitarity triangle}
\end{figure}
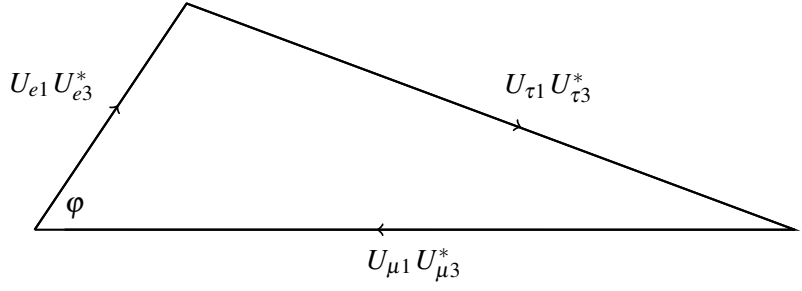
Let us analyse an example of a Majorana triangle, which is shown in figure~\ref{fig:Majtri}. This triangle is a typical triangle that depends only on Majorana phases. The Majorana phases give the directions of the sides of the Majorana unitary triangles. The arrows in figure~\ref{fig:Majtri} have no meaning! They can be reversed if one makes, for example the rephasing:
\begin{equation}
\nu_3\,\longrightarrow\,\nu_3' \,=\, -\nu_3\,.
\end{equation}
The angle $\varphi$ is a Dirac-type phase, since:
\begin{equation}
\varphi\,=\,\arg\left(-U_{e1}\,U^{\ast}_{e3}\,U^{\ast}_{\tau1}\,U_{\tau3}\right)\,=\,\pi-(\gamma_3-\gamma_1)\,.
\end{equation}
Majorana triangles provide necessary and sufficient conditions for having CP invariance with Majorana neutrinos, namely:
\begin{itemize}
\item Vanishing of their common area:
\begin{equation}
A\,=\,\frac12\left|\imag Q\right|\,;
\end{equation}
\item Orientation of all the ``collapsed'' triangles along the real axis or the imaginary axis. If one of these triangles $T_{ik}$ is parallel to the imaginary axis, that means that the neutrinos $i,k$ have opposite CP parities.
\end{itemize}
The $0\nu\beta\beta$, which is sensitive to Majorana-type phases, is proportional to the quantity $m_{ee}$ given by
\begin{equation}
\begin{split}
m_{ee}^2 &\,=\,
m_1^2\,|U_{e1}|^4 \,+\, m_2^2\,|U_{e2}|^4 \,+\, m_3^2\,|U_{e3}|^4 \\
&\,+\, 2\,m_1\,m_2\,|U_{e1}|^2\,|U_{e1}|^2 \cos2\beta_1 
\,+\, 2\,m_1\,m_3\,|U_{e1}|^2\,|U_{e3}|^2 \cos 2\gamma_1\\
&\,+\, 2\,m_2\,m_3\,|U_{e2}|^2\,|U_{e3}|^2 \cos2(\beta_1\,-\,\gamma_1)\,. 
\end{split}
\end{equation}
Note that the angle $(\beta_1\,-\,\gamma_1)$ is the argument of $\left(U^{\ast}_{e1}\,U_{e2}\,U_{e1}\,U^{\ast}_{e3}\right)$ which is not a rephasing invariant Dirac-type quartet. If one adopts the parametrisation given by eq.~\eqref{eq:spn1} with the definition given in eq.~\eqref{eq:spn}, one then has:
\begin{equation}
m_{ee}\,=\,\left|c_{13}^2\left(m_1\,c_{12}^2\,+m_2\,e^{-i\alpha_1}\,s_{12}^2 \right)
\,+\,m_3\,e^{-i\alpha_2}\,s_{13}^2\right|\,.
\end{equation}
This is the reason why this parametrisation is useful for the analysis of $0\nu\beta\beta$.

One may ask whether the effective neutrino mass matrix can be determining from experiment. We have seen that in the WB where the charged lepton mass matrix is diagonal, real one has:
\begin{equation}
m_{\ell}\,=\,\diag(m_e,\,m_{\mu},\,m_{\tau})\,,\qquad 
m_{\nu}\,=\,\begin{pmatrix}
m_{ee}  & m_{e\mu} & m_{e\tau}\\
m_{e\mu}  & m_{\mu\mu} & m_{\mu\tau}\\
m_{e\tau}  & m_{\mu\tau} & m_{\tau\tau}\,.
\end{pmatrix}\,.
\end{equation} 
One can use rephasing freedom to make $m_{ee}$, $m_{\mu\mu}$,
$m_{\tau\tau}$ real, but $m_{e\mu}$, $m_{e\tau}$, $m_{\mu\tau}$
complex. Altogether we end up with 6 real parameters and three phases,
a total of 9 real parameters. Let us then compare with the physical
quantities in $m_{\nu}$, that can be obtained trough feasible
experiments: two mass squared differences, $\Delta m_{12}^2$, $\Delta
m_{23}^2$, three mixing angles $\theta_{12}$, $\theta_{23}$,
$\theta_{13}$, the imaginary of the quartet $\imag Q$ from CP
violation in $\nu$ oscillations and $0\nu\beta\beta$ from the
measurement of $m_{ee}$. This is a total of 7 measurable quantities
(Glashow's counting). We arrive at the dreadful conclusion ($7<9$)
that no presently conceivable set of feasible experiments can fully
determine the effective neutrino matrix~\cite{Frampton:2002yf}. Some
of the ways out:
\begin{itemize}
\item Postulate ``texture zeroes'' in $m_{\nu}$~\cite{Frampton:2002yf}: various sets of zeroes allowed by experiment.
\item Postulate $\det (m_{\nu})=0$. This leads to 7 parameters in $m_{\nu}$~\cite{Branco:2002ie}.
\end{itemize}
Let us now derive CP-odd Weak-basis invariants in the leptonic sector with Majorana neutrinos. The relevant part of the Lagrangian is:
\begin{equation}
\Lag_{\text{mass}}\,=\,-\overline{\ell}_L\,m_{\ell}\,\ell_R \,-\,
\frac12\nu_L^{\mathsf{T}}\,C\,m_{\nu}\,\nu_L \,+\,
\frac{g}{\sqrt2}\overline{\ell}_L\gamma_{\mu}\,\nu_L\,W^{\mu}
\,+\,\text{H.c.}\,.
\end{equation}
The CP transformation properties of the various fields are dictated by the part of the Lagrangian which conserves CP, namely the gauge interactions. One has to keep in mind the fact that the gauge sector of the SM does not distinguish the various families of  fermions. The most general CP transformations which leave $\Lag_{\rm{gauge}}$ invariant is:
\begin{subequations}
\begin{align}
\rm{CP}\,\ell_L\,(\rm{CP})^{\dagger} &\,=\, W_L\,\gamma^0\,C\,\overline{\ell}_L^{\mathsf{T}}\,,\\
\rm{CP}\,\nu_L\,(\rm{CP})^{\dagger} &\,=\, W_L\,\gamma^0\,C\,\overline{\nu}_L^{\mathsf{T}}
\,,\\
\rm{CP}\,\ell_R\,(\rm{CP})^{\dagger} &\,=\, W_R\,\gamma^0\,C\,\overline{\ell}_R^{\mathsf{T}}
\end{align}
\end{subequations}
where $W_L$, $W_R$ are unitarity matrices acting in generation space. The Lagrangian of the leptonic sector conserves CP if and only if the leptonic mass matrices $m_{\nu}$, $m_{\ell}$ satisfy 
\begin{equation}
W_L^{\mathsf{T}}\,m_{\nu}\,W_L\,=-\,m^{\ast}_{\nu}\,,
\qquad
W_L^{\dagger}\,m_{e}\,W_R\,=\,m^{\ast}_{e}\,,
\end{equation}
or
\begin{equation}
W_L^{\dagger}\,\tilde{h}_{\nu}\,W_L\,=\,\tilde{h}^{\ast}_{\nu}\,,
\qquad
W_L^{\dagger}\,h_{e}\,W_L\,=\,h^{\ast}_{e}\,,
\end{equation}
where $\tilde{h}_{\nu}\equiv h^{\ast}_{\nu}\,$, $h_{\nu}\equiv m_{\nu}\,m^{\ast}_{\nu}\,$ and $h_{e}\equiv m_{e}\,m^{\dagger}_{e}\,$. One then gets 
\begin{equation}
W_L^{\dagger}\,\left[\tilde{h}_{\nu},\,h_{\ell}\right]\,W_L 
\,=\, \left[\tilde{h}^{\ast}_{\nu},\,h_{\ell}^{\ast}\right]
\,=\, \left[\tilde{h}^{\mathsf{T}}_{\nu},\,h_{\ell}^{\mathsf{T}}\right]
\,=\, -\left[\tilde{h}_{\nu},\,h_{\ell}\right]^{\mathsf{T}}\,.
\end{equation} 
Making the cube of both sides, one gets
\begin{equation}
W_L^{\dagger}\,\left[\tilde{h}_{\nu},\,h_{\ell}\right]^3\,W_L 
\,=\, -\left(\left[\tilde{h}_{\nu},\,h_{\ell}\right]^3\right)^{\mathsf{T}}\,.
\end{equation} 
Therefore, CP invariance implies 
\begin{equation}
\tr \left[\tilde{h}_{\nu},\,h_{\ell}\right]^3\,=\,0\,,
\end{equation} 
which is valid for an arbitrary number of generations. This relation, first derived by~\cite{Bernabeu:1986fc}, can be written for three generations in terms of measurable quantities as
\begin{equation}
\tr \left[\tilde{h}_{\nu},\,h_{\ell}\right]^3\,=\,-6\,i\,(m_{\mu}^2-m_e^2)\,(m_{\tau}^2-m_{\mu}^2)\,
(m_{\tau}^2-m_e)\,\Delta m^2_{21}\,\Delta m^2_{31}\,\Delta m^2_{32}\,\imag Q\,,
\end{equation}
where 
\begin{equation}
\imag Q\,=\,\frac18\, \sin2\theta_{12}\,\sin2\theta_{13}\,\sin2\theta_{23}\,sin\delta\,.
\end{equation}
This invariant is sensitive to Dirac type CP violation. For three generations, the vanishing of this invariant is the necessary and sufficient for the absence of Dirac-type CP violation.
Using the previous method, one can derive invariants sensitive to Majorana-type CP violation~\cite{Branco:1986gr}. The following invariant is sensitive to Majorana-type CP violation:
\begin{equation}
I^{\rm{CP}}_{\rm{Majorana}}\,=\,\imag\tr\left(
m_{\ell}\,m_{\ell}^{\dagger}\,m_{\nu}^{\ast}\,m_{\nu}\,m_{\nu}^{\ast}\,m_{\ell}^{\mathsf{T}}\,m^{\ast}_{\ell}\,m_{\nu}
\right)\,.
\end{equation}
The simplest way to check that $I^{\rm{CP}}_{\rm{Majorana}}$ is sensitive to Majorana-type CP violation is by evaluating it in the case of two generations of Majorana neutrinos 
\begin{equation}
I^{\rm{CP}}_{\rm{Majorana}}\,=\,\frac14\,m_1\,m_2\,\Delta m^2_{21}\,(m_{\nu}^2-m_e^2)\,\sin^2\theta\sin2\gamma\,,
\end{equation}
where the PMNS matrix $U$ is parametrised as
\begin{equation}
U=\begin{pmatrix}
\cos\theta & -\sin\theta e^{i\gamma}\\
\sin\theta e^{-i\gamma} & \cos\theta
\end{pmatrix}\,.
\end{equation}
The invariant $I^{\rm{CP}}_{\rm{Majorana}}$ vanishes, for example, if $\gamma=\pi/2\,$, since this corresponds to having CP invariance, with the two neutrinos with opposite CP parities.


\subsection{Violating $3\times3$ CKM Unitarity}

Suppose that one drops the requirement of $3\times3$ unitarity. How many parameters are there in the $3\times3$ CKM matrix? By taking into account the elements of the CKM matrix $V$:
\begin{equation}
\begin{pmatrix}
\ V_{ud}\  & \ V_{us}\  & \ V_{ub}\  & \cdots \\
V_{cd} & V_{cs}   & V_{cb}  & \cdots   \\
V_{td} & V_{ts}   &  V_{tb}   & \cdots \\
\vdots & \vdots &  \vdots  & \ddots
\end{pmatrix}\,.
\end{equation}
One counts 9 moduli plus 4 $(9-5)$ rephasing invariant phases for a total of 13 parameters. A convenient choice for 4 independent rephasing invariant phases is:
\begin{subequations}
\begin{align}
\beta&\,\equiv\,\arg\left(-V_{cd}\,V_{tb}\,V_{cb}^{\ast}\,V_{td}^{\ast}\right)\,,\\
\gamma&\,\equiv\,\arg\left(-V_{ud}\,V_{cb}\,V_{ub}^{\ast}\,V_{cd}^{\ast}\right)\,,\\
\chi\,=\,\beta_s&\,\equiv\,\arg\left(-V_{cb}\,V_{ts}\,V_{cs}^{\ast}\,V_{tb}^{\ast}\right)\,,\\
\chi'&\,\equiv\,\arg\left(-V_{us}\,V_{cd}\,V_{ud}^{\ast}\,V_{cs}^{\ast}\right)\,,
\end{align}
\end{subequations}
The SM with three generations predicts a series of exact relations among the 13 measurable (in principle) quantities. Again we should emphasise  that the relation
\begin{equation}
\alpha \,+\, \beta \,+\, \gamma \,=\, \pi\,,
\end{equation}
is not a test of unitarity. It is true, by definition!
\begin{subequations}
\begin{align}
\alpha&\,\equiv\,\arg\left(-V_{td}\,V_{ub}\,V_{ud}^{\ast}\,V_{tb}^{\ast}\right)\,,\\
\beta&\,\equiv\,\arg\left(-V_{cd}\,V_{tb}\,V_{cb}^{\ast}\,V_{td}^{\ast}\right)\,,\\
\gamma&\,\equiv\,\arg\left(-V_{ud}\,V_{cb}\,V_{ub}^{\ast}\,V_{cd}^{\ast}\right)\,.
\end{align}
\end{subequations}

In the derivation of the unitary relations, it is useful to adopt a convenient phase convention~\cite{Botella:2005fc}. Without loss of generality one can choose:
\begin{equation}
\arg(V)\,=\,\begin{pmatrix}
\ 0\  & \chi' & \ -\gamma  \\
\pi & 0  & 0  \\
-\beta & \pi+\chi   & 0  
\end{pmatrix}\,.
\end{equation}
We have used the 5 rephasing degrees of freedom to fix 5 of the nines phases. We are left with 4 phases.

In the case of the SM where the CKM matrix is strictly unitary, one has exact relations predict by the SM, such as~\cite{Branco:2008ai,Botella:2002fr}:
\begin{subequations}
\begin{align}
|V_{ub}| &\,=\, \frac{|V_{cd}| |V_{cb}|}{ |V_{ub}| }\,\frac{\sin\beta}{\sin(\beta+\gamma)}\,,\\
\sin\chi &\,=\, \frac{|V_{td}| }{ |V_{ts}| }\,\frac{|V_{cd}| }{|V_{cs}| }\,\sin\beta\,,\\
\frac{|V_{ub}|}{|V_{tb}|} &\,=\, \frac{\sin\beta}{\sin\gamma}\,\frac{|V_{tb}|}{ |V_{ud}| }\,,\\
\sin\chi &\,=\, \frac{|V_{us}| |V_{ub}|}{ |V_{cs}|  |V_{cb}| }\,\sin(-\chi+\chi'+\gamma)\,.
\end{align}
\end{subequations}
Violation of any of these exact relations signals the presence of New Physics which may involve deviations of $3\times3$ unitarity or not. The presence of New Physics contributions to $B_d-\bar{B}_d$ and $B_s-\bar{B}_s$ mixings affects the extraction of $|V_{td}|$, $|V_{ts}|$ from the data, even in the framework of New Physics which respects $3\times3$ unitarity. An example of that is the Supersymmetric extension of the SM. 
In many of the extensions of the SM, the dominant effect of New Physics arises from new contributions to  $B_d-\bar{B}_d$ and $B_s-\bar{B}_s$ mixings, which is convenient to parametrise as:
\begin{equation}
M^q_{12}\,=\,\left(M^q_{12}\right)^{\rm{SM}}\,r_q^2\,e^{2i\theta_q}\,,
\end{equation}
with $q=d,s$. Thus the mass difference $\Delta M_{B_d}$ is now given by	
\begin{equation}
\Delta M_{B_d}\,=\,r_d^2\,(\Delta M_{B_d})^{\rm{SM}}\,,
\end{equation}
that affects the extraction of $|V_{td}|$ from experiment. On the other hand, the mass difference $\Delta M_{B_s}$ is 	
\begin{equation}
\Delta M_{B_s}\,=\,r_s^2\,(\Delta M_{B_s})^{\rm{SM}}\,,
\end{equation}
that affects the extraction of $|V_{ts}|$.

From the CP asymmetries $S_{J/\psi\,K_s}$ and $S_{\rho^+\,\rho^-}$ given by
\begin{subequations}
\begin{align}
S_{J/\psi\,K_s}&\,=\,\sin(2\beta+2\theta_d)\,\equiv\,\sin2\bar{\beta}\,,\\
S_{\rho^+\,\rho^-}&\,=\,\sin(2\alpha-2\theta_s)\,\equiv\,\sin2\bar{\alpha}\,.
\end{align}
\end{subequations}
How to detect the presence of New Physics? The answer is: one can use
the exact relations predicted by the SM. The extraction $\theta_d$
from
\begin{equation}
|V_{ub}| \,=\, \frac{|V_{cd}| |V_{cb}|}{ |V_{ub}| }\,\frac{\sin\beta}{\sin(\beta+\gamma)}\,,
\end{equation}
which then leads to 
\begin{equation}
  \tan\theta_d\,=\, \frac{R_u\,\sin(\gamma+\bar{\beta})-\sin\bar{\beta}}{\cos\bar{\beta}\,-\,R_u\,\cos(\gamma\,+\,\bar{\beta})}\,,
\end{equation}
where
\begin{equation}
R_u\,=\,\frac{|V_{ud}|\,|V_{ub}|}{|V_{cd}|\,|V_{cb}|}\,. 
\end{equation}
While to extract $\theta_s$ one must use the exact relation
\begin{equation}
\sin\chi\,=\, \frac{|V_{us}| |V_{ub}|}{ |V_{cs}|  |V_{cb}| }\,\sin(\gamma-\chi+\chi')\,,
\end{equation}
which then leads to 
\begin{equation}
\tan\theta_s\,=\, \frac{\sin\bar{\chi}\,-\,C\,\sin(\gamma\,-\,\bar{\chi})}{C\,\cos(\gamma-\bar{\chi})-\cos\bar{\chi}}\,,
\end{equation}
where
\begin{equation}
C\,=\,\frac{|V_{us}|\,|V_{ub}|}{|V_{cs}|\,|V_{cb}|}\,. 
\end{equation}
To an excellent approximation one has~\cite{Silva:1996ih}:
\begin{equation}
\sin\chi\,=\,\frac{|V_{us}|^2}{|V_{ud}|^2}\,\frac{\sin\beta\,\sin\gamma}{\sin{\gamma+\beta}}\,,
\end{equation}
or~\cite{Botella:2005fc}:
\begin{equation}
\sin\chi\,=\,\frac{|V_{td}|}{|V_{ts}|}\,\frac{|V_{cd}|}{|V_{cs}|}\sin\beta\,.
\end{equation}
If either $(\gamma,\chi)$ or $(\frac{\Delta M_{B_d}}{\Delta
  M_{B_s}},\chi)$ are measured with some precision, one has novel
stringent tests of the SM, where contribution of New Physics can be
significant. At this point, the following point should be emphasised. There is clear evidence for a complex CKM matrix even if one allows for the presence of New Physics~\cite{Botella:2005fc}. This is essentially due to the evidence for a non-vanishing $\gamma$, which is not contaminated by the presence of New Physics!

Since we are considering experimental tests of $3\times3$ unitarity of the CKM matrix, one should ask the following questions:
\begin{itemize}
\item Can one have self-consistent extensions of the SM, where deviations of $3\times3$ unitarity of the CKM matrix may occur?
\item Can these deviations be naturally small?
\end{itemize}
The answer to both questions is positive! In the next section we describe an extension of the SM with the addition of a vector-like quarks.

  
\subsection{SM with the addition of a isosinglet down-type quark}

We shall consider in this section extensions of the SM with
vector-like isosinglet quarks of $Q=-1/3$ and $Q=2/3$. One question
one may raise the question whether the addition of vector-like quarks
to SM can bring important features to solve flavour issues presented
in the SM. We point out several reasons to consider vector-like
quarks:
\begin{enumerate}
\item they provide a self-consistent framework with naturally small
  violations of $3\times3$ unitarity of the CKM matrix.
\item Lead to naturally small Flavour Changing Neutral Currents (FCNC) mediated by $Z_{\mu}$.
\item Provide the simplest framework to have spontaneous CP violation~\cite{Bento:1990wv,Bento:1991ez},
  with a vacuum Phase generating a non-trivial CKM phase. An important
  requirement is that there is experimental evidence of a complex CKM
  matrix even if one allows for the presence of New Physics.
\item Provide New Physics contributions to $B_d-\bar{B}_d$ mixing and
  $B_s-\bar{B}_s$ mixing.
\item Provide a simple solution to the \emph{Strong CP problem}, which
  does not require Axions.
\item May contribute to the understanding of the observed pattern of
  fermion masses and mixing.
\item Provide a framework where there is a common origin of all CP violations~\cite{Branco:2003rt}:
\begin{itemize}
\item[(i)] CP violation in the Quark Sector;
\item[(ii)] CP violation in the Lepton Sector detectable through neutrino oscillations $U_{e3}\neq0$ and ``relatively large''. This is a great feature!
\item[(iii)]  CP violation need to generate the Baryon Asymmetry of the Universe through Leptogenesis.
\end{itemize}
\end{enumerate}
There is nothing strange in having deviations of $3\times3$
unitarity. The PMNS matrix in the leptonic sector in the context of
type-one seesaw ($\nu$SM) is not $3\times3$ unitarity.

For simplicity let us study the Minimal Model where one adds a
vector-like quark field $D^0$ into SM. This down-type quark particle
$D^0$ has the property that both chiral fields $D^0_L$ and $D^0_R$ are
$SU(2)_L$ singlets with electric charge $Q=-1/3$ (one could also have
introduced a isosinglet of the up-type instead with electric charge
$Q=2/3$). To complete the fermionic content of this Minimal Model we
introduce 3 right-handed neutrinos $\nu^0_{Rj}$. The Higgs sector is
just extended with a neutral complex singlet field $S$.

Since we want to have \emph{Spontaneous CP Violation}, we impose CP
invariance at the Lagrangian level, i.e. all couplings are taken
real. We add a $Z_4$ symmetry, under which the SM fields transform as~\cite{Bento:1990wv,Bento:1991ez,Branco:2003rt}:
\begin{equation}
\ell^0\,\rightarrow\,i\,\ell^0\,,\quad
e_{Rj}^0\,\rightarrow\,i\,e^0_{Rj}\,,\quad
\nu_{Rj}^0\,\rightarrow\,i\,\nu^0_{Rj}\,,
\end{equation}
and the remaining SM fermions transform trivially. The new particles
transform as
\begin{equation}
D^0\,\rightarrow\,-\,D^0\,,\quad
S\,\rightarrow\,-\,S\,.
\end{equation}
The discrete symmetry $Z_4$ is crucial to obtain a solution of the
Strong CP problem and Leptogenesis. The scalar potential contains
various terms which do not have phase dependence but there are terms
with phase dependence, which are given by
\begin{equation}
V_{\rm{phase}}(\phi,S)\,=\,\left[\mu^2\,+\,\lambda_1\,S^{\ast}S+\lambda_2\phi^{\dagger}\phi\right](S^2\,+\,{S^{\ast}}^2)\,+\,
\lambda_3\,(S^4\,+\,{S^{\ast}}^4)\,.
\end{equation}
There is a range of the parameters of the Higgs potential, where the minimum is at:
\begin{equation}
\langle\phi\rangle\,=\,\frac{v}{\sqrt2}\,,\qquad
\langle S\rangle\,=\,\frac{V}{\sqrt2}\,e^{i\theta}\,.
\end{equation}
The most general $SU(2)_C\times SU(2)_L\times U(1)\times Z_4$ invariant Yukawa couplings 
\begin{equation}
-\Lag\,=(\bar{u}^0 \; \bar{d}^0)_{Li}\,\left[ g_{ij}\,\phi\,d^0_{Rj} \,+\, h_{ij}\,\phi\,u^0_{Rj} \right]
\,+\, \overline{M}\;\overline{D}^0_L\,D_R^0\,+\,\text{H.c.}\,.
\end{equation}
This implies that the quark mass matrix  for down-type quarks has the following form:
\begin{equation}
\label{eq:Md}
\overline{\mathbf{d}}_L\,\mathcal{M}\, \mathbf{d}_R\,=\,\begin{pmatrix}
\bar{d}^0_{1L} &\bar{d}^0_{2L} &\bar{d}^0_{3L} &\overline{D}^0_{L}
\end{pmatrix}
\,
\begin{pmatrix}
\\
m_d & 0\\
\\
\begin{matrix}
M_1 & M_2 & M_3
\end{matrix}
& \overline{M}
\end{pmatrix}
\,
\begin{pmatrix}
d^0_{1R} \\ d^0_{2R} \\  d^0_{3R} & \\ D^0_{R}
\end{pmatrix}\,,
\end{equation}
where
\begin{equation}
\label{eq:Mj}
M_j \,=\, f_j\,V\,e^{i\theta} \,+\,  f^{\prime}_j\,V\,e^{i\theta}\,.
\end{equation}
The  \emph{zero} $3\times$ column in the down-type quark matrix in eq.~\eqref{eq:Md} is due to the $Z_4$ symmetry. The down quarks masses are then obtained through the diagonalisation:
\begin{equation}
\mathcal{U}^{\dagger}_L\,\mathcal{M}\mathcal{M}^{\dagger}\,\mathcal{U}_L \,=\, \diag(m_d^2,m_s^2,m_b^2,m_D^2)\,.
\end{equation}
Defining the block-entries of the unitary matrix $\mathcal{U}$ as
\begin{equation}
\mathcal{U}_L\,=\,\begin{pmatrix}K & R \\ S & T\end{pmatrix}\,,
\end{equation}
one can easily derive approximative the effective down quark mass matrix $m_{\rm{eff}}\,m^{\dagger}_{\rm{eft}}$ as:
\begin{equation}
K^{-1}\,m_{\rm{eff}}\,m_{\rm{eft}}^{\dagger}\,K\,=\,\diag(m_d^2,m_s^2,m_b^2)\,,
\end{equation}
where $m_{\rm{eff}}\,m_{\rm{eft}}^{\dagger}$ is given by
\begin{equation}
m_{\rm{eff}}\,m_{\rm{eft}}^{\dagger}\,=\,m_d\,m_d^{\dagger}\,-\,\frac{m_d\,M^{\dagger}\,M\,m_d^{\dagger}}{M\,M^{\dagger}\,+\,\overline{M}^2}\,.
\end{equation}
It is worth to point out that $m_{\rm{eff}}\,m^{\dagger}_{\rm{eff}}$ is complex since the combination $M^{\dagger}\,M$ is complex because of eq.~\eqref{eq:Mj}.
A remarkable feature of the Model is that the phase $\theta$ arising from $\langle S\rangle$, generates a non-trivial CKM phase, provided $|M_j|$ and  $\overline{M}$ are of the same order of magnitude, which it is natural. We shall now see that within this model that deviations of $3\times3$ unitarity and Flavour-Changing Neutral Currents are naturally small. Let us write down the charged and neutral currents in the model:
\begin{equation}
\Lag_{W} \,=\, -\frac{g}{\sqrt2}\,
\begin{pmatrix} \bar{u} & \bar{c} & \bar{t} \end{pmatrix}_L
\gamma^{\mu}\,
\begin{pmatrix} d \\ s \\ b \\ D \end{pmatrix}_L\,
W_{\mu}^{+}\,,
\end{equation}
and
\begin{equation}
\begin{split}
\Lag_{Z} \,=\, -\frac{g}{2\cos\theta_W}\,&\left\{
\begin{pmatrix} \bar{u} & \bar{c} & \bar{t} \end{pmatrix}_L\,
\gamma^{\mu}\,
\begin{pmatrix} u \\ c \\ t \end{pmatrix}_L
\,+\,
\begin{pmatrix} \bar{d} & \bar{s} & \bar{b} & \overline{D} \end{pmatrix}
\begin{pmatrix} 
K^{\dagger}\,K & &K^{\dagger}\,R\\ \\
R^{\dagger}\,K & &R^{\dagger}\,R
\end{pmatrix}\,
\gamma^{\mu}\,
\begin{pmatrix} d \\ s \\ b \\ D \end{pmatrix}
\right.\\
&\left.\phantom{\begin{pmatrix} d \\ s \\ b \\ D \end{pmatrix}}\,-\,\sin^2\theta_W\,J^{\mu}_{\rm{em}}
\right\}\,
Z_{\mu}\,,
\end{split}
\end{equation}
Let us  now quantify the deviations of $3\times3$ unitarity. Since the full $6\times6$ matrix $\mathcal{U}_L$ is unitary,  the following relations are verified:
\begin{equation}
\label{eq:KRST}
K^{\dagger}\,K\,+\,S^{\dagger}\,S \,=\, \mathbf{1}\,,\qquad
R^{\dagger}\,R\,+\,T^{\dagger}\,T \,=\, \mathbf{1}\,,\qquad
R^{\dagger}\,K\,+\,T^{\dagger}\,S \,=\, \mathbf{0}\,,
\end{equation}
and one derives
\begin{equation}
S\,\approx\,-\frac{M\,m_d^{\dagger}\,K}{\overline{M}^2}\,\rightarrow\mathcal{O}\left(\frac{m}{M}\right)\,,
\end{equation}
and therefore, making use of the relations given in eq.~\eqref{eq:KRST}, one concludes that the deviations of $3\times3$ unitarity,
\begin{equation}
K^{\dagger}\,K\,=\,\mathbf{1}\,-\,\mathcal{O}\left(\frac{m^2}{M^2}\right)
\end{equation}
are naturally small. Note that there is nothing strange about violations of $3\times3$ unitarity. The PMNS matrix is not unitary in the framework of seesaw mechanism, type-one.
  
Can extensions of the SM with vector-like quarks ``solve'' some of the tensions between SM and experiment? The answer is yes! In the framework of an extension of the SM, with one $Q=2/3$ vector like quark,  it has been shown that the tensions can be solved and various correlations are predicted~\cite{Botella:2012ju}. But, the important point is for experiment/theory to confirm that deviations are really there. In table~\ref{tab:utfit} we present deviations between experiment and theory presented from UTfit Collaboration in the  ICHEP2012 Conference.

\begin{table}
\caption{\label{tab:utfit} Deviations between experiment and theory. From UTfit Collaboration, Cecilia Tarantino talk at ICHEP2012 Conference.}
\begin{center}
\begin{tabular}{cccc}
\hline\hline
& \textbf{Prediction} & \textbf{Measurement} & \textbf{Pull}\\
\hline
$\sin2\beta$ & $0.81\pm0.05$ & $0.680\pm0.023$ & $2.4$\\ 
$\gamma$ & $68^{\circ}\pm3^{\circ}$ & $76^{\circ}\pm11^{\circ}$ & $<1$\\ 
$\alpha$ & $88^{\circ}\pm4^{\circ}$ & $91^{\circ}\pm6^{\circ}$ & $<1$\\ 
$|V_{cb}|\cdot10^3$ & $42.3\pm0.9$ & $41.0\pm1.0$ & $<1$\\ 
$|V_{ub}|\cdot10^3$ & $3.62\pm0.14$ & $3.82\pm0.56$ & $<1$\\ 
$\epsilon_K\cdot10^3$ & $1.96\pm0.20$ & $2.23\pm0.01$ & $1.4$\\ 
$BR(B\rightarrow\tau\nu)\cdot10^4$ & $0.82\pm0.08$ & $1.67\pm0.30$ & $-2.7$\\
\hline\hline
\end{tabular}
\end{center}
\end{table}

\subsubsection*{Leptonic Sector}

We recall that the leptonic fields transform under $Z_4$ as:
\begin{equation}
\ell^0\,\rightarrow\,i\,\ell^0\,,\quad
e_{Rj}^0\,\rightarrow\,i\,e^0_{Rj}\,,\quad
\nu_{Rj}^0\,\rightarrow\,i\,\nu^0_{Rj}\,.
\end{equation}
This constrains the Yukawa Lagrangian terms as:
\begin{equation}
-\Lag_{\ell} \,=\,
\bar{\ell}^0_L\,G_{\ell}\,\phi\,e^0_R \,+\,
\bar{\ell}^0_L\,G_{\nu}\,\phi\,\nu^0_R \,+\,
\frac12\,{\nu^0_R}^{\mathsf{T}}\,C\,\left(f_{\nu}\,S\,+\,f'_{\nu}S^{\ast}\right)\,\nu^0_R
\,+\, \text{H.c.}\,,
\end{equation}
which  after spontaneous breaking become as
\begin{equation}
-\Lag_{\ell} \,=\,
\bar{e}^0_L\,G_{\ell}\,\phi\,e^0_R \,+\,
\frac12\,n_L^{\mathsf{T}}\,C\,\mathcal{M}^{\ast}\,n_L
\,+\, \text{H.c.}\,,
\end{equation}
with
\begin{equation}
n_L\,\equiv\,
\begin{pmatrix}
\nu^0_L \\  (\nu_R)^C
\end{pmatrix}\,.
\end{equation}
The $6\times6$ matrix is then given by  
\begin{equation}
\mathcal{M}_{\nu} \,=\,
\begin{pmatrix}
0 & m \\
m^{\mathsf{T}} & M_{\nu}
\end{pmatrix}\,,
\end{equation}   
where  
\begin{equation}
m_{\ell}\,=\,\frac{v}{\sqrt2}\,G_{\ell}\,,\qquad
m\,=\,\frac{v}{\sqrt2}\,G_{\nu}\,,
\end{equation}
and $M_{\nu}$ is given by
\begin{equation}
M_{\nu} \,=\, \frac{V}{\sqrt2}\,(f^{+}_{\nu}\cos\theta \,+\, if^{-}_{\nu}\sin\theta)\,,
\end{equation}
where $f^{\pm}\equiv f_{\nu}\pm f_{\nu}\,$.

In the weak-basis where $m_{\ell}$ is diagonal, real, the light
neutrino masses and low energy leptonic mixing are obtained from
\begin{equation}
K^{\dagger}\left(m\,M_{\nu}^{-1}\,m^{\mathsf{T}} \right)\,K\,=\, \diag(m_1,m_2,m_3)\,
\end{equation}
where $m$ is real, but since $M_{\nu}$ is a generic complex matrix,
the effective light neutrino mass $m_{\nu}$ is also a generic complex
symmetric matrix. Thus, the matrix $K$ has three complexes phases: one
Dirac-type and two Majorana-type.
 
\subsubsection*{Conclusions}

Vector-like quarks provide a very interesting scenario for New
Physics.  They are a crucial ingredient in the simplest realistic
model of spontaneous CP violation where a complex CKM matrix is
generated from a vacuum phase. They provide a consistent framework
where there are naturally small deviations of $3\times3$ unitarity in
the CKM matrix, leading to naturally small Z-FCNC. They provide a
simple solution to the \emph{Strong CP problem}, without the need of
introducing \emph{axions}.

The Standard Model and its CKM mechanism for mixing and CP violation
is in good agreement with experiment. This is a remarkable fact in
view of the \emph{large amount of data}: $|V_{us}|$, $|V_{ub}|$,
$|V_{cb}|$, $\gamma$ completely fix the CKM matrix. Then with no free
parameters, one has to accommodate a large number of measurable
quantities like $\varepsilon_K$, $B_d-\overline{B}_d$ mixing,
$B_s-\overline{B}_s$, $\beta$, $\beta_s$, rare $B$-decays, rare kaon
decays, etc., etc. Unfortunately there are \emph{hadronic
  uncertainties}.
 
There is room for New Physics which could be detected in LHCb and
future super-B factories. The spectrum of Fermion Masses and the
Pattern of quark and lepton mixing remains one of the Fundamental
Questions in Particle Physics. It is very likely that detectable New
Physics be involved in the solution of the Flavour Puzzle. The
observation of neutrino oscillation is a strong indication to search
for an extension of the SM that can account for neutrino masses.


\subsection{Baryon Asymmetry of the Universe}

We now address the question how to generate the Baryon Asymmetry of
the Universe (BAU). The ingredients to dynamically generate BAU from
an initial state with zero Baryon Asymmetry were formulate in 1967 by
Sakharov~\cite{Sakharov:1967dj} :
\begin{itemize}
\item[(i)] Baryon number violation;
\item[(ii)] C and CP violation;
\item[(iii)] Departure from thermal equilibrium.
\end{itemize}
All these ingredients exist in the SM, but it has been established
that in the SM, one cannot generate the observed BAU:
\begin{equation}
\eta_B\,\equiv\,\frac{n_B\,-\,n_{\bar{B}}}{n_{\gamma}}\,=\,(6.20\pm0.15)\times10 ^{-10}\,,
\end{equation}
where $n_B$, $n_{\bar{B}}$, $n_{\gamma}$ correspond to number densities of baryons, anti-baryons and photons at present time, respectively. There are mainly two reasons why the SM cannot generate sufficient BAU:
\begin{itemize}
\item[(i)] CP violation in the SM is too small. Indeed one has:
\begin{equation}
\frac{\tr\left[H_u,H_d\right]^3}{T^{12}_{\rm{EW}}}\,\simeq\,10^{-20}\,.
\end{equation}
\item[(ii)]Successful Baryogenesis requires a strongly first order phase transition which would require a light Higgs mass:
\begin{equation}
m_H\,\leq\,70\,\text{GeV}\,.
\end{equation}
\end{itemize}
One concludes that an explanation of the observed BAU requires New
Physics beyond the SM. Leptogenesis, suggested by Fukugita and
Yanagida is one the simplest and most attractive mechanism to generate
BAU.  In the framework of leptogenesis, BAU is generated through out of equilibrium decays of
right-handed neutrinos that create a lepton asymmetry which is in turn
converted into a baryon asymmetry by $(B+L)$ violating but $(B-L)$
conserving sphaleron interactions.

The $\nu$SM, i.e. the extension of the SM consisting of adding 3
right-handed neutrinos has all the ingredients to have Leptogenesis. For recent reviews, see~\cite{Davidson:2008bu,Branco:2011zb}.
\begin{equation}
\begin{split}
-\Lag_m & \,=\,
\bar{\nu}^0_L\,m_D\,\nu^0_R \,+\,
\frac12\,{\nu^0_R}^{\mathsf{T}}\,C\,M_R\,\nu^0_R \,+\,
\bar{e}^0_L\,m_{\ell}\,e^0_R \,+\,
\text{H.c.} \\
& \,=\,
\frac12\,n_L^{\mathsf{T}}\,C\,\mathcal{M}^{\ast}\,n_L \,+\, 
\bar{e}^0_L\,m_{\ell}\,e^0_R \,+\,
\text{H.c.}\,,
\end{split}
\end{equation}
with 
\begin{equation}
n_L\,=\,
\begin{pmatrix}
\nu_L^0\\
(\nu_R^0)^c
\end{pmatrix}
\,.
\end{equation}
The full neutrino mass matrix is a $6\times6$ matrix:
\begin{equation}
\mathcal{M}\,=\,
\begin{pmatrix}
0 & m \\
m^{\mathsf{T}} & M_R
\end{pmatrix}\,,
\end{equation}
diagonalised by 
\begin{equation}
V^{\mathsf{T}}\,\mathcal{M}^{\ast}\,V \,=\,
\begin{pmatrix}
d & 0\\ 0 & D
\end{pmatrix}
\,,
\end{equation}
with 
\begin{equation}
d\equiv\diag(m1,m2,m3) \qquad \text{and} \qquad
D\equiv\diag(M1,M2,M3)\,. 
\end{equation}
The unitary matrix $V$ can be described by
\begin{equation}
V\,=\, \begin{pmatrix}
K & G\\
S & T
\end{pmatrix}\,.
\end{equation}
One can show that:
\begin{equation}
S^{\dagger}\,\approx\,K^{\dagger}\, m\, M^{-1}_R\,, \qquad \text{and} \qquad
G^{\dagger}\,\approx\,K^{\dagger}\, m\,T^{\ast}\,D^{-1}\,\approx\,m\,D^{-1}\,,
\end{equation}
so that one gets the usual seesaw formula
\begin{equation}
-K^{\dagger}\,m\,\frac{1}{M}\,m^{\mathsf{T}}\,K^{\ast} \,=\, d\,.
\end{equation}
The leptonic charged current interactions are:
\begin{equation}
-\frac{g}{\sqrt2}\,(\bar{\ell}_{iL}\,\gamma_{\mu}\,K_{ij}\,\nu_{jL}
\,+\, \bar{\ell}_{iL}\,\gamma_{\mu}\,G_{ij}\,N_{jL})\,W^{\mu}
\,+\,\text{H.c.}\,.
\end{equation}
Let us count the parameters in the leptonic sector. Without loss of
generality, one can choose a Weak basis where the charged lepton mass
matrix is diagonal, real and also the right-handed neutrino mass
matrix is diagonal, real. In this basis, the Yukawa coupling matrix
$Y_D$ entering in the Dirac neutrino mass matrix is an arbitrary
complex matrix. Moreover, 3 of 9 phases in $Y_D$ can be eliminated by
rephasing. So altogether one has 3 charged lepton masses, 3 diagonal
entries in $M_R$, 9 real parameters in $Y_D$ plus 6 phases in
$Y_D$. This gives a total of 21 parameters.

The \emph{Lepton-Asymmetry} generated through CP violating decays of
the Heavy neutrinos,
\begin{equation}
N\,\longrightarrow\,\ell\,+\,H\,,
\end{equation}
within unflavoured Leptogenesis approximation, which do not include
flavour effects, is given by
\begin{equation}
A^{j} \,=\, \frac{\sum_i\,\Gamma^j_i\,-\,\overline{\Gamma}_i^j}{\sum_i\,\Gamma^j_i\,+\,\overline{\Gamma}_i^j}
\,\propto\,
\sum_{k\neq j}\,C_K\,\imag\left[(m_D^{\dagger}\,m_D)^2_{jk}\right]
\end{equation}
where
\begin{equation}
\Gamma^j_i\equiv\Gamma(N_j\,\longrightarrow\,\ell_i\,+\,H)\,,\qquad
\overline{\Gamma}^j_i\equiv\Gamma(N_j\,\longrightarrow\,\bar{\ell}_i\,+\,H)\,.
\end{equation}
Taking into account the Casas and Ibarra parametrisation~\cite{Casas:2001sr}:
\begin{equation}
m_D\,=\,i\,U_{\nu}\,\sqrt{d}\,R\,\sqrt{D}\,,
\end{equation}
with $R$ a  complex orthogonal matrix,
one sees that the combination $m_D^{\dagger}\,m_D$ is given by:
\begin{equation}
m_D^{\dagger}\,m_D\,=\,-\,\sqrt{D}\,R^{\dagger}\,d\,R\,\sqrt{D}\,,
\end{equation}
and thus one concludes that leptogenesis is independent of $U_{\nu}$.

In general, it is not possible to establish a connection between CP
asymmetries needed for leptogenesis and CP violation detectable in
Neutrino Oscillations. \emph{One may have leptogenesis even if
  $U_{\nu}$ is real}~\cite{Rebelo:2002wj}. The connection
may be established with further theoretical
assumptions~\cite{Branco:2002xf,Frampton:2002qc,Branco:2002kt}.

Can on have a WB invariant which is sensitive to the CP violating
phases entering in \emph{Unflavoured Leptogenesis}? It is indeed
possible. The WB invariant sensitive to the CP violating phases
entering in \emph{Unflavoured Leptogenesis} is given
by~\cite{Branco:2001pq}:
\begin{equation}
\begin{split}
I&\,\equiv\, \imag\tr\left[h\,H\,M_R^{\ast}\,h^{\ast}\,M_R\right]
\,=\,
M_1\,M_2\,(M_2^2-M_1^2)\,\imag(h_{12}^2)\\
&
\,+\,
M_1\,M_3\,(M_3^2-M_1^2)\,\imag(h_{13}^2)\,+\,
M_2\,M_3\,(M_3^2-M_2^2)\,\imag(h_{23}^2)\,,
\end{split}
\end{equation}
where $h\equiv m_D^{\dagger}\,m_D$ and $H\equiv M^{\dagger}_R\,M_R\,$.

\section{Conclusions}

Neutrino Oscillations provide clear evidence for Physics beyond the SM
and the discovery of $U_{e3}\neq0$ opens up the exciting possibility
of detecting leptonic Dirac-type CP violation through neutrino
oscillations.

Leptogenesis is an attractive framework to generate BAU which can
occur in the framework of $\nu$.SM It is difficult to test
experimentally, but one should try to find a framework where this is
possible.

It is urgent to conceive \emph{feasible experiments} which can measure
physical quantities in $m_{\nu}$ beyond the seven quantities mentioned
by Glashow et al. Difficult but what looks impossible today, may be
possible tomorrow! It would be very nice if some years from now, we
have a workshop with a title like:
\begin{center}
\textit{\textbf{\Large The leptonic unitarity triangle fit}}
\end{center}  

\section*{Acknowledgments}
We would like to thank the organisers of the Third IDPASC School,
specially Carlos Merino, for the very warm hospitality extended to us
and for the very nice atmosphere in the School. This work was
supported by \emph{Funda\c{c}\~ao para a Ci\^encia e a Tecnologia}
(FCT, Portugal) through the projects CERN/FP/123580/2011, PTDC/FISNUC/
0548/2012 and CFTP-FCT Unit 777 (PEst-OE/FIS/UI0777/2013) which are
partially funded through POCTI (FEDER). The work of D.E.C.~was also
supported by Associa\c c\~ao do Instituto Superior T\'ecnico para a
Investiga\c c\~ao e Desenvolvimento (IST-ID).

\addcontentsline{toc}{section}{References} 
\bibliographystyle{JHEP}
\bibliography{refs}

\end{document}